\documentclass[aip, chaos, reprint, twocolumn, groupedaddress, showpacs, floatfix]{revtex4-1}

\usepackage{lipsum}
\usepackage[english]{babel}
\usepackage[utf8]{inputenc}
\usepackage{hyperref}
\usepackage{verbatim}
\usepackage{enumerate}
\usepackage{url}
\usepackage{times}
\usepackage{tikz}
\newcommand{\twocol}[1]{#1}

\usepackage{xcolor}
\definecolor{dgreen}{RGB}{0,148,0}
\definecolor{dred}{rgb}{0.5,0,0}
\definecolor{purple}{rgb}{0.5,0,0.5}
\definecolor{darkblue}{RGB}{0,0,140}
\definecolor{darkgreen}{RGB}{0,148,0}
\hypersetup{colorlinks,%
citecolor=darkblue,%
filecolor=green,%
linkcolor=blue,%
urlcolor=blue,%
}

\usepackage{makeidx}
\usepackage{overpic}

\newcommand{\sdSN}{black curve}
\newcommand{\sdHB}{blue curve}
\newcommand{\sdHC}{purple curve}
\newcommand{\sdBT}{purple dot}
\newcommand{\sdCP}{black dot}
\newcommand{\sdGH}{blue dot}
\newcommand{\sdSLH}{red dot}
\newcommand{\sdCPC}{green dot}
\newcommand{\sdSNLC}{black dashed curve}
\newcommand{\sddr}{white region} 
\newcommand{\sdL}{light gray shaded region} 
\newcommand{\sdLL}{dark gray shaded region} 
\newcommand{\sdLdr}{yellow region} 
\newcommand{\sdLCRL}{blue region} 
\newcommand{\bdSADDLE}{black dashed curve}
\newcommand{\bdSTABLENODE}{black curve}

\newcommand{\bdSTABLEFOCUS}{red line}
\newcommand{\bdUNSTABLEFOCUS}{red dashed curve}
\newcommand{\bdUNSTABLELC}{blue dashed curve}
\newcommand{\bdSN}{black dot}
\newcommand{\bdHB}{blue dot}
\newcommand{\bdHC}{purple dot}
\newcommand{\bdPD}{green dashed line}
\newcommand{\bdOscExtr}{blue} 
\newcommand{\bdOscExtrCoexist}{magenta} 
\newcommand{\dIndicLine}{green line} 
\newcommand{\dIndicRect}{green rectangle} 

\newcommand{\lblSN}[2]{\put(#1,#2){\textcolor{black}{SN}}}
\newcommand{\lblHB}[2]{\put(#1,#2){\textcolor{blue}{HB}}}
\newcommand{\lblHC}[2]{\put(#1,#2){\textcolor{purple}{HC}}}

\newcommand{\lblBTone}[2]{\put(#1,#2){\textcolor{purple}{BT$_1$}}}
\newcommand{\lblBTtwo}[2]{\put(#1,#2){\textcolor{purple}{BT$_2$}}}
\newcommand{\lblBTthree}[2]{\put(#1,#2){\textcolor{purple}{BT$_3$}}}
\newcommand{\lblCP}[2]{\put(#1,#2){\textcolor{black}{CP}}}
\newcommand{\lblGH}[2]{\put(#1,#2){\textcolor{blue}{GH}}}
\newcommand{\lblCPC}[2]{\put(#1,#2){\textcolor{dgreen}{CPC}}}
\newcommand{\lblSLH}[2]{\put(#1,#2){\textcolor{red}{SLH}}}

\newcommand{\lbldr}{D} 
\newcommand{\lblL}{L} 
\newcommand{\lblLL}{2L} 
\newcommand{\lblLdr}{D/L} 
\newcommand{\lblmixedosc}{MO}
\newcommand{\lbllibration}{LB}
\newcommand{\lblpartial}{PS}
\newcommand{\lblAP}{AP}
\newcommand{\lblLibrationAndMixedOsc}{\lbllibration/\lblmixedosc}
\newcommand{\lblLockLibrationAndMixedOsc}{\lblL/\lbllibration/\lblmixedosc}
\newcommand{\lblLockLibrationAndMixedOscnoLC}{\lblL/\lbllibration/\lblmixedosc/ w/o LC}

\newcommand{\axlblhalf}[2]{ 
\put(60,0){#1}
\put(0,47){#2}
}
\newcommand{\axlblquarter}[2]{ 
\put(60,-3){#1}
\put(-5,62){#2}
}
\newcommand{\pnlblhalf}[1]{\put(0,70){(#1)}} 
\newcommand{\pnlblquarter}[1]{\put(-3,70){(#1)}} 
\newcommand{\pnhdln}[1]{\put(13,70){#1}}
\newcommand{\insetaxtick}[3]{\put(#1,#2){\tiny\textcolor{dgreen}{#3}}}
\newcommand{\pointertootherfig}[3]{\put(#1,#2){\footnotesize\textcolor{dgreen}{#3}}}
\newcommand{\lbltickTwoDBif}{
\axlblquarter{$\omega$}{$\kappa$}
\put(0,16){$0$}
\put(0,55){$2$}
\put(21,0){$-1$}
\put(48,0){$0$}
\put(73,0){$1$}
}
\newcommand{\lbltickTwoDBifPhi}{
\axlblquarter{$\omega$}{$\phi$}
\put(0,35){$0$}
\put(-4,5){$-\pi$}
\put(0,67){$\pi$}
\put(21,0){$-1$}
\put(48,0){$0$}
\put(73,0){$1$}
}
\newcommand{\lblPhiKapMultimodeDriftChaos}{
\axlblquarter{$\phi$}{$\kappa_{12}$}
\put(4,0){$0$}
\put(49,0){$\pi$}
\put(90,0){$2\pi$}
\put(-4,19){$1.5$}
\put(-4,55){$2.5$}
}

\usepackage{amsmath, amssymb}
\usepackage{bm}

\newcommand{\R}{\mathbb{R}}

\newcommand{\T}{\mathbb{T}}

\newcommand{\Z}{\mathbb{Z}}

\newcommand{\tz}{\rightarrow0} 
\def\txtd{{\textnormal{d}}}
\newcommand{\tr}{{\rm tr}}

\renewcommand{\d}{\text{d}}
\newcommand{\dotxp}[1]{\frac{\txtd #1}{\txtd t}} 
\renewcommand{\epsilon}{\varepsilon}
\newcommand{\maxLE}{\lambda_{\text{max}}}

\newcommand{\A}{\mathcal{A}}
\newcommand{\ada}{\A}

\newcommand{\highlight}[1]{#1}

\begin{document}

\preprint{AIP/123-QED}

\title{Complex dynamics in adaptive phase oscillator networks}
\author{Benjamin J\"{u}ttner${}^{a}$}%
\author{Erik A.~Martens${}^{b,c}$}%
\email{erik.martens@math.lth.se}

\affiliation{%
\twocol{\mbox}
{${}^a$Department of Applied Mathematics and Computer Science, Technical
University of Denmark, 2800 Kgs.~Lyngby, Denmark}\\
\twocol{\mbox}
{${}^b$Centre for Mathematical Sciences, Lund University, Sölvegatan 18B, 221 00 Lund, Sweden}\\
{${}^c$Center for Translational Neurosciences, University of Copenhagen,
Blegdamsvej 3, 2200 Copenhagen, Denmark}\\
}
\date{\today}

\begin{abstract}%
Networks of coupled dynamical units give rise to collective dynamics such as the synchronization of oscillators or neurons in the brain.
The ability of the network to adapt coupling strengths between units in accordance with their activity arises naturally in a variety of contexts, including neural plasticity in the brain, and adds an additional layer of complexity: the dynamics on the nodes influence the dynamics of the network and vice versa.
We study a minimal model of Kuramoto phase oscillators including a general adaptive learning rule with three parameters (strength of adaptivity, adaptivity offset, adaptivity shift), mimicking \highlight{learning paradigms based on spike-time-dependent-plasticity (STDP)}. Importantly, the strength of adaptivity allows to tune the system away 
away from the limit of the classical Kuramoto model, corresponding to stationary coupling strengths and no adaptation, and thus, to systematically study the impact of adaptivity on the collective dynamics.
We carry out a detailed bifurcation analysis for the minimal model consisting of $N=2$ oscillators. \highlight{The non-adaptive Kuramoto model exhibits very simple dynamic behavior, drift or frequency-locking; but once the strength of adaptivity exceeds a critical threshold non-trivial bifurcation structures unravel}: \highlight{A symmetric adaptation rule results in  multi-stability and bifurcation scenarios, and an asymmetric adaptation rule generates even more intriguing and rich dynamics, including a period doubling-cascade to chaos as well as oscillations displaying features of both librations and rotations simultaneously. Generally, adaptation improves the synchronizability of the oscillators. Finally, we also numerically investigate a larger system consisting of $N=50$ oscillators and compare the resulting dynamics with the case of $N=2$ oscillators.}
\end{abstract}

\pacs{05.45.-a, 05.45.Gg, 05.45.Xt, 02.30.Yy}
\keywords{adaptive Kuramoto model} 
\maketitle

\begin{quotation}
Synchronization is a ubiquitous phenomenon manifesting itself in a range of natural and technological systems~\cite{Strogatz2003,PikovskyBook2001}. The presence or absence of synchronization orchestrates the proper functioning of complex networks, such as in  neural networks in the brain~\cite{Singer1995,Uhlhaas2006} or power transmission networks~\cite{Rohden2012}.
A paradigmatic model to study synchronization is Kuramoto's model that describes the dynamics of phase oscillators. Many variants have been studied in literature~\cite{Acebron2005,BickMartens2020}; here, we are concerned with the dynamics that emerges when coupling strengths adapt according to the oscillator activity~\cite{Seliger2002plasticity,Maistrenko2007multistability}. Such ability to adapt the coupling strength has been receiving much attention lately~\cite{Berner2019,Berner2020birth,Lucken2016noise,Ruangkriengsin2022low,Thamizharasan2022} and finds applications for models of synaptic plasticity and learning in the brain~\cite{Gerstner1996}.
We consider the adaptive Kuramoto model with two oscillators in the limit of stationary coupling and investigate how the strength of adaptivity affects the dynamics of the network and find that generally, the synchronizability increases with a larger level of adaptivity. Nontrivial bifurcations, unknown to the Kuramoto model with stationary coupling, emerge at a critical adaptivity threshold. We analytically and numerically determine these bifurcations and their stability boundaries for several types of learning paradigms. Finally, numerical simulations give a glimpse into how the dynamics observed for small systems with two oscillators carries over to larger systems with a larger number of oscillators.
\end{quotation}

\section{Introduction}
The synchronization of coupled oscillators is a fascinating manifestation of self-organization  --- indeed, self-emergent synchronization is a central process to a spectacular range of natural and technological systems, including the beating of the
heart~\cite{Michaels1987},
flashing fireflies~\cite{Buck1968},
pedestrians on a bridge locking their gait~\cite{Strogatz2005a},
genetic clocks~\cite{Danino2010},
pendulum clocks hanging on a beam~\cite{Huygens1967ab}, mechanical oscillators~\cite{MartensThutupalli2013,Goldsztein2022}
superconducting Josephson junctions~\cite{Wiesenfeld1998},
chemical oscillations~\cite{Taylor2009,Calugaru2020},
metabolic oscillations in yeast cells~\cite{Dano1999},
and life cycles of phytoplankton~\cite{Massie2010}, and networks of neurons in the brain~\cite{Singer1995,Uhlhaas2006,YamakouHjorthMartens2020}

A desirable property in real-world oscillator networks is the presence of synchronization whenever it ensures the proper functioning of the network: in the realm of technology, for instance, the AC current between generators and consumers in a power grid need to stay synchronized to ensure ideal power transmission and ultimately avoid power blackouts~\cite{Rohden2012}; and wireless networks require synchronized clocks~\cite{klinglmayr2012guaranteeing} to ensure safe data transmission; in the realm of biology, the proper functioning of the heart requires that the rhythmic electric activation of cardiac cells stays coordinated~\cite{Michaels1987}.

While the network interaction between dynamic units may give rise to intriguing collective behaviors such as synchronization,
adding the ability to adapt the (coupling) weights on a network according to the dynamics on its nodes leads to  co-evolutionary network dynamics ~\cite{GrossBlasius2008}.
Indeed --- ``Intelligence is the ability to adapt to change''~\cite{WashingtonPost} --- the adaptive dynamics of co-evolutionary networks may allow to increase their functional robustness~\cite{Strogatz2000,Gkogkas2022}. 
Adaptive co-evolutionary networks appear in a wide range of systems, including the vascular network~\cite{MartensKlemm2017,MartensKlemm2018}, the glymphatic network of the brain~\cite{Mestre2020,Kelley2022}, osteocyte network formation~\cite{Taylor2017}, social networks~\cite{Skyrms2009dynamic}, and, in particular, in neural networks in the brain where neural plasticity plays an important role for learning~\cite{Gerstner1996,Gerstner2002}, but also for the progression of certain neuro-degenerative diseases~\cite{Goriely2020neuronal}.
In the current context, this raises the question if adaptivity in a coupled oscillator network can increase its ability to synchronize.

We devise a special variation of the Kuramoto model with adaptive coupling weights. While previous studies have considered such systems from various perspectives, we here introduce a parameterization that allows to systematically deviate from the traditional Kuramoto system with stationary uniform coupling and systematically study the effects of adaptation and the related bifurcation behavior.

This article is structured as follows.
In Sec.~\ref{sec:model}, we explain our adaptive Kuramoto model and how we parameterize it.
In Sec.~\ref{sec:analysis}, we consider $N=2$ oscillators and carry out a detailed bifurcation analysis for several important parameter cases:~\ref{sec:Dyn1D} Non-adaptive limit (classical Kuramoto model);~\ref{sec:Dyn2D} 
\highlight{Symmetric adaptation ($\alpha=\beta=0$);~\ref{sec:Dyn3D} Asymmetric adaptation.}
In Sec.~\ref{sec:N50} we numerically investigate larger oscillator systems with $N=50$ and compare the resulting dynamic behavior with the results for $N=50$.
Sec.~\ref{sec:discuss} concludes with a summary and discussion of our results.

\section{Model}\label{sec:model}
\subsection{General model}\label{sec:generalModel}
We consider a general model of $N$ oscillators $l\in[N]:=\{1,\ldots,N\} $ with adaptive coupling. The oscillator phase $\phi_l=\phi_l(t)\in\T=\R/2\pi\Z$ then evolves according to
\begin{align}
\dotxp{\phi_l} &= \omega_l + \frac{1}{N}\sum_{m=1}^N \kappa_{lm}\;g{(\phi_m- \phi_l) }\
\end{align}
with intrinsic frequencies $\omega_l$. Oscillators $l$ and $m$ interact via the interaction function $g$ and are coupled with the time-dependent coupling strength $\kappa_{lm}\in\R$. The coupling strengths evolve according to
\begin{align}
    \dotxp{\kappa_{lm}} &= \epsilon(\ada(\phi_l-\phi_m) - \kappa_{lm})
\end{align}
where the adaptation or learning rule $\ada=\ada(\phi)$ defines how the coupling adapts according to the oscillator phases (states); i.e., the adaptation/learning rule is defined via a local interaction between oscillator's phases. The second term with $\kappa_{lm}$ guarantees boundedness of the coupling strengths. The time scale at which adaptation may occur is set by $\epsilon$: oscillator phases evolve on time-scales $\sim 1$ and coupling adapts on a time scale $\sim 1/\epsilon$.

\paragraph{Coupling function and adaptation rule.} Specifically, we consider the simplest version of such a model and suppose that $g$ and $\ada$ are periodic functions in $\phi$. \highlight {We may think of these functions as truncations of Fourier series to first order}, i.e.,
the coupling function becomes
\begin{align}
    g(\phi)=\sin{(\phi+\alpha)} ,\
\end{align}
and the adaptation rule (or function) becomes
\begin{align}\label{eq:adaptationrule}
 \ada(\phi) &= a_0 + a_1 \cos{(\phi+\beta)},\
\end{align}
\highlight{where $\alpha$ and $\beta$ tune the level of cosine versus sine interaction in the coupling function and adaptation rule, respectively.}
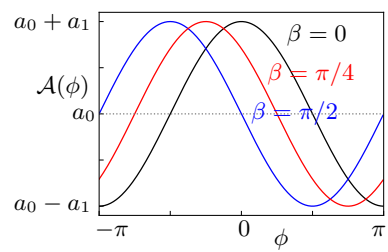
\begin{figure}[htp!]
\begin{overpic}[width=.5\linewidth]{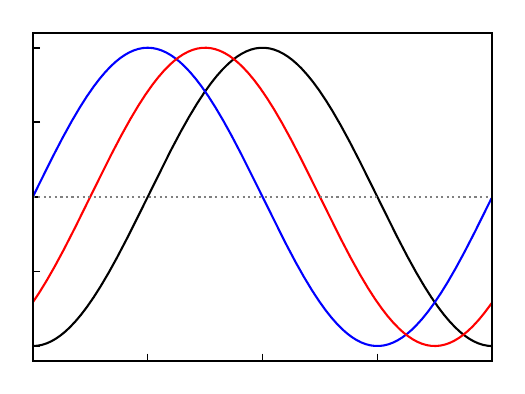}
  \put(-13,44){$\ada(\phi)$}
  \put(60,-3){$\phi$}
  \put(-2,36){$a_0$}
  \put(-20,64){$a_0+a_1$}
  \put(-20,8){$a_0-a_1$}
  \put(49,0){$0$}
  \put(90,0){$\pi$}
  \put(5,0){$-\pi$}
  \put(64,60){\textcolor{black}{$\beta=0$}}
  \put(58,48){\textcolor{red}{$\beta=\pi/4$}}
  \put(53,36){\textcolor{blue}{$\beta=\pi/2$}}
\end{overpic}
 \caption{
 The adaptation rules $\ada(\phi)$ for distinct values of adaptation shifts $\beta$ (shown for $a_1>0$).
 }
 \label{fig:our_adas}
\end{figure}
Thus, the oscillator dynamics is defined by the Kuramoto-Sakaguchi model with phase-lag $\alpha$; the adaptation/learning rule has an \emph{adaptation offset}, $a_0\neq 0$, and the \emph{adaptivity} $a_1$ (i.e., the strength of adaptation). 
\highlight{The adaptation rule is inspired by models of spike-time-dependent plasticity, a Hebbian synaptic learning rule~\cite{Caporale2008}, and can be approximated in phase oscillator networks~\cite{Duchet2022} as we do here. The shape of the adaptation rule induces different types adaptation or learning~\cite{Maistrenko2007multistability,Lucken2016noise,Berner2019}}. In this context, the \emph{adaptation shift} $\beta$ tunes the type of interaction. Restricting our attention to positive adaptivity, $a_1>0$, the following adaptation regimes may be distinguished (see also Fig.~\ref{fig:our_adas}):
\begin{enumerate}
 \item \highlight{Symmetric adaptation with in-phase gain} ($\beta=0$)
 amplifies (or suppresses) the coupling strength between oscillators that have phase difference close to 0 (or $\pi$).

 \item \highlight{Symmetric adaptation with anti-phase gain} ($\beta=\pi$)
 amplifies (or suppresses) the coupling strength between oscillators that have phase difference close to  $\pi$ (or 0).

 \item \highlight{Non-symmetric adaptation} ($\beta\neq 0,\pi$)
 amplifies (or suppresses) coupling strengths between oscillators with non-trivial phase difference; e.g., for the odd adaptation function  with $\beta=\pi/2$, coupling strengths associated with phase difference close to $-\pi/2$ (or $\pi/2$) are amplified (or suppressed).

\end{enumerate}

\highlight{
Note that the system has the parameter symmetry $(a,\beta) \mapsto (-a,\beta+\pi)$. Thus, if we consider $\beta=0$ for both positive ($a>0$) and negative ($a<0$) adaptivity, we need not include the case of $\beta=\pi$ in our investigation.
In addition to these special cases we are also interested in investigated mixed-type learning rules such as the case $\beta=\pi/4$, see Sec.~\ref{sec:beta0_25pi}.
}

\highlight{Symmetric/asymmetric adaptation corresponds to an even/odd adaptation function; as we shall see further below, the respective types of adaptation lead to symmetric/asymmetric coupling, i.e., undirected/directed network weights.}

\paragraph{Reduced/rescaled model and its symmetries.} Furthermore, we note that the 'self-couplings', $\kappa_{ll}(t)$, are decoupled from the oscillator phases; vice versa, if $g(0)=\sin{\alpha}=0$, oscillators are independent from the self-coupling. But even if $\sin{\alpha}\neq 0$, the post-transient or asymptotic values $\kappa_{ll}(\infty)\rightarrow a_0 + a_1 \cos{\beta}$ can be absorbed into the intrinsic frequencies,
$\omega_l + \highlight{\sin{\alpha}\cdot}\kappa_{ll}(\infty)\mapsto \omega_l$. We may therefore disregard all coupling terms $\kappa_{ll}$
\footnote{Note that this property is due to the model's special form that is invariant with regards to stretching time; other models, e.g., with intrinsic dynamics of the form \unexpanded{$\dot{\phi}_l=f(\phi_l) + \sum_l  g(\phi_l,\phi_m)$} do not necessarily share this property, an example being the Theta neuron model.}.
Furthermore, we may reduce the number of parameters involved by appropriately rescaling parameters and variables,
$$a_0t \mapsto t,\quad
\frac{\omega_l}{a_0} \mapsto  \omega_l,\quad
\frac{\kappa_{lm}}{a_0}\mapsto \kappa_{lm}, \quad
\frac{\epsilon}{a_0}\mapsto\epsilon.
$$
Since $\epsilon>0$, we restrict ourselves to $a_0 > 0$. Introducing the \emph{(relative) adaptivity}, $a:=a_1/a_0$, we then obtain the governing equations for the model:
\begin{subequations}\label{eq:reducedmodel}
\begin{align}
    \dotxp{\phi_l} &= \omega_l + \frac{1}{N}\sum_{m=1}^N \kappa_{lm}\;\sin{(\phi_m- \phi_l+\alpha)},\\
    \dotxp{\kappa_{lm}} &= \epsilon(1+a\cos{(\phi_l-\phi_m+\beta)} - \kappa_{lm}),\
\end{align}
\end{subequations}
where $l\neq m$.

Since  $g(\phi)$ and $\ada(\phi)$ only depend on phase differences, \eqref{eq:reducedmodel} is invariant to shifts in constant phase and frequency, i.e., $\phi_l(t) \mapsto \phi_l(t)+\Phi+\Omega t$ with constants $\Phi,\Omega\in\R$.

\paragraph{Non-adaptive case ($a=0$).} 
When $a=0$, the model asymptotically strives to $\kappa_{lm}=1$ for all $l,m\in[N]$, corresponding to the Kuramoto model with (rescaled) uniform stationary coupling strength, $\kappa=1$. The emergent synchronization in the model can be characterized by the order parameter
\begin{align}\label{eq:orderparameter}
 Z(t) &=\frac{1}{N} \sum_{l=1}^N e^{i\phi_l(t)}.\
\end{align}
The Kuramoto model exhibits the following asymptotic behavior~\cite{Kuramoto1984,Strogatz2000,BickMartens2020}. In the continuum limit, $N\to\infty$, when the coupling strength is subcritical ($\kappa<\kappa_c$), the order parameter converges to 0, $|Z|\rightarrow 0$, corresponding to incoherent oscillations; vice versa, for supercritical coupling ($\kappa>\kappa_c$), the order parameter approaches a non-zero value corresponding to partial synchronization, $|Z|\rightarrow C>0$. \highlight{(Note that for finite oscillator systems, the order parameter is subject to pseudo-random fluctuations and it is preferable to consider the time-averaged order parameter, $\langle |Z(t)| \rangle_t$. In the subcritical regime, fluctuations will prevent the order parameter  from precisely attaining 0.)} 
For the continuum limit $N\rightarrow \infty$, the threshold for this transition  occurs at $\kappa_c=2/(\pi g(0))$~\cite{Strogatz2000,Gkogkas2022},
where $g(\omega)$ is a unimodal frequency distribution with maximum $g(0)$ and width $\sigma$. Note that we normalized the coupling to $1$; thus, the synchronization threshold, $\sigma_c$, is indirectly determined  via $1=2/(\pi g(0))$.
Vice versa, if $a\neq 0$, we allow the network to be adaptive and we move away from the manifold corresponding to the Kuramoto model with stationary coupling.
\highlight{We will also use the order parameter \eqref{eq:orderparameter} to characterize the adaptive dynamics for $N=50$ in Sec.~\ref{sec:N50}.}

\subsection{Two oscillator model}\label{sec:Ntwomodel}
We consider the minimal network with $N=2$ oscillators.
Introducing $\phi := \phi_1-\phi_2$ and $\omega:=\omega_1-\omega_2$, the governing equations become
\begin{subequations}\label{eq:governingRescal}
\begin{align}
\dotxp{\phi} &= \omega + \frac{1}{2}\kappa_{12}\sin(\alpha-\phi) - \frac{1}{2}\kappa_{21}\sin(\alpha+\phi),\label{eq:governingRescalPhi}\\
\dotxp{\kappa_{12}} &= \epsilon(1 + a \cos(\beta+\phi)-\kappa_{12}),\label{eq:governingRescalKap12}\\
\dotxp{\kappa_{21}} &= \epsilon(1 + a \cos(\beta-\phi)-\kappa_{21}).\label{eq:governingRescalKap21}
\end{align}
\end{subequations}
This system of equations has the additional symmetries
\begin{subequations}
\begin{align}
  (a,\beta) &\mapsto (-a,\beta+\pi)
  \label{eq:UsefulSymm_abeta}\\
  (a,\alpha,\phi) &\mapsto (-a,\alpha+\pi,\phi+\pi)
  \label{eq:UsefulSymm_aalphaphi}\\
  (\alpha,\beta,\kappa_{12},\kappa_{21}) &\mapsto (-\alpha,-\beta,\kappa_{21},\kappa_{12}),\
  \label{eq:UsefulSymm_minusalphabeta}
\end{align}
\end{subequations}
\highlight{Sometimes it is useful to consider the dynamics in terms of the  difference $\Delta:=\kappa_{12}-\kappa_{21}$ and sum $\Sigma:=\kappa_{12}+\kappa_{21}$. Eqs.~\eqref{eq:governingRescalKap12} and \eqref{eq:governingRescalKap21} then become
\begin{subequations}\label{eq:sumdiff}
  \begin{align}
    \dotxp{\Delta} &= \epsilon(a[\cos(\beta+\phi)-\cos(\beta-\phi)]-\Delta),\\
    \dotxp{\Sigma} &= \epsilon(2+a[\cos(\beta+\phi)+\cos(\beta-\phi)]-\Sigma).
  \end{align}
\end{subequations}
}

Due to \eqref{eq:UsefulSymm_abeta}, we may restrict the parameter range to $-\pi/2<\beta\leq \pi/2$;
due to \eqref{eq:UsefulSymm_aalphaphi}, we may restrict the parameter range to $-\pi/2<\alpha\leq \pi/2$; furthermore, \eqref{eq:UsefulSymm_minusalphabeta} allows to restrict either the range of $\beta$ or $\alpha$ to positive values, and we chose to restrict $0 \leq \beta\leq \pi/2$. However, observe that
all symmetry transformations also affect other parameters and variables; e.g.,
the symmetry \eqref{eq:UsefulSymm_aalphaphi} effectively swaps (near-)phase-locked states where $\phi$ is close to zero, with (near-)antiphase states where $\phi$ is close to $\pi$.
Thus, if a phase-locked stationary state is stable for $\alpha=0$ with adaptivity $a$, then an antiphasic stationary state is also stable for $\alpha=\pi$ with $-a$.
There are several other symmetries, see Appendix~\ref{app:symm}\highlight{, some of which are reflected in the stability diagrams (Figs.~\ref{fig:2Dstab},~\ref{fig:stabbeta0_5pi_alpham0_1pi},~\ref{fig:stabbeta0_25pi_alpha0_1pi}, and~\ref{fig:alpham0_1pi_beta0_25pi_stabbif}(a)).}

Furthermore, we consider $\epsilon>0$.
As our analysis will show, we find interesting nontrivial dynamic behavior (at least)
if $\epsilon=0.2$. We use this value throughout the analysis unless specified otherwise.

\section{Analysis}\label{sec:analysis}
Certain parameter choices lead to effectively lower-dimensional dynamics. It is instructive to first consider these cases (Secs.~\ref{sec:Dyn1D} and \ref{sec:Dyn2D}) before analyzing the more general case leading to three dimensional dynamics (Sec.~\ref{sec:Dyn3D}).

\subsection{Non-adaptive limit (classical Kuramoto model)}\label{sec:Dyn1D}
Non-adaptive dynamics is obtained for two limiting cases: either $\epsilon = 0$, so that $\kappa_{12}$ and $\kappa_{21}$ are constants; or the adaptivity is zero, $a=0$, and the coupling strengths asymptotically become identical with $\kappa_{12}(t),\kappa_{21}(t)\rightarrow 1$ as $t\rightarrow\infty$, irrespective of the value of $\beta$. In either cases, the system is effectively one-dimensional with the dynamics given by ~\eqref{eq:governingRescalPhi}, i.e., the system corresponds to the classical Kuramoto model. Since we are only interested in $\epsilon>0$ we consider the case of $a=0$ and let $\kappa_{12}=\kappa_{21}=1$ to consider the post-transient dynamics.
Eq.~\eqref{eq:governingRescalPhi} can be cast as
\begin{align}\label{eq:governing1D}
\dotxp{\phi} = \omega - \cos\alpha\sin\phi.
\end{align}
Since we restricted $\alpha\in(-\pi/2,\pi/2]$, we have $\cos\alpha\geq0$. Excluding the effectively decoupled case with $\cos\alpha\neq0$, we may rescale time and $\omega$ with $\cos\alpha$ to obtain
\begin{align}\label{eq:governing1DRescal}
\dotxp{\phi} = \omega - \sin\phi.
\end{align}
When the frequency mismatch exceeds the coupling strength, $|\omega|>1$,  there are no fixed points
and the distance $\phi=\phi_1-\phi_2$ keeps increasing, amounting to \emph{drifting} oscillators.
When the frequency mismatch is smaller, $|\omega|<1$, we have the two equilibria
\begin{align}
\phi_- := \arcsin\omega,\,\,\,\,\,\phi_+ := \pi-\arcsin\omega,\label{eq:EQ1D}
\end{align}
$\phi_-$ is a stable equilibrium while $\phi_+$ is unstable (note that stability types are inverted for $\cos\alpha<0$).
These equilibria correspond to (frequency-)locked states with the two oscillators having constant phase difference.
When the oscillators share their intrinsic frequencies, $\omega=0$ , the oscillators are phase-locked.
For $|\omega|=1$, the two equilibria collide in a saddle-node bifurcation on an invariant cycle (SNIC).
These drift/locked states, as well as the saddle-node bifurcation, remain as a basic dynamic structure when the system is adaptive with $a\neq 0$.

\subsection{\highlight{Symmetric adaptation} ($\beta=0,\pi$)}\label{sec:Dyn2D}

\subsubsection{Reduction}
The adaptation rule is symmetric (or an even function) for $\beta=0$ and $\beta=\pi$, so that Eqs.~\eqref{eq:governingRescalKap12} and \eqref{eq:governingRescalKap21} attain identical structure. As a result, the three dimensional dynamics of \eqref{eq:governingRescal} is asymptotically described by dynamics on a two dimensional subspace \highlight{with symmetric coupling, $\kappa_{12}=\kappa_{21}$, corresponding to undirected network weights}. To see this,
we substitute $\beta=0$ into Eqs.~\eqref{eq:sumdiff} which simplifies to
\begin{subequations}\label{eq:sumdiffbeta0}
  \begin{align}
    \dotxp{\Delta} &= -\epsilon\Delta\label{eq:sumdiffbeta0diff},\\
    \dotxp{\Sigma} &= \epsilon(2[1+a\cos\phi]-\Sigma)\label{eq:sumdiffbeta0sum}.
  \end{align}
\end{subequations}
Defining $\kappa(t):=\kappa_{12}(t)=\kappa_{21}(t)=\Sigma(t)/2$,
\eqref{eq:governingRescalPhi} and \eqref{eq:sumdiffbeta0sum}
become
\begin{subequations}\label{eq:governing2D}
\begin{align}
\dotxp{\phi} &= \omega -\kappa\cos\alpha\sin\phi  \label{eq:governing2DPhi},\\
\dotxp{\kappa} &= \epsilon (1+a\cos\phi  - \kappa)\label{eq:governing2DKap}.\
\end{align}
\end{subequations}
This system describes the dynamics on a
two-dimensional subspace, i.e. the plane defined by $\kappa_{12}=\kappa_{21}$ in the full $(\phi,\kappa_{12},\kappa_{21})$ phase space. As is evident from inspecting~\eqref{eq:sumdiffbeta0diff}, this plane is globally attracting since $\Delta \to 0 $ as $t \to \infty$. 

Eqs.~\eqref{eq:governing2D} admit for two degenerate cases: (i) if $\omega =\cos\alpha=0$, $\phi$ is constant and thus $\kappa(t)\to1+a\cos\phi$ exponentially fast; (ii) if $\cos\alpha=0$ with $\omega\neq0$, we have $\phi(t)=\phi_0+\omega t$ so that oscillators drift apart and $\kappa(t)$ forever undergoes an oscillation.

Since we restricted $\alpha \in (-\pi/2,\pi/2]$, we may assume $\cos\alpha\geq0$. Further restricting $\cos\alpha\neq0$ we may rescale  time and related parameters with $\cos\alpha$ to obtain
\begin{subequations}\label{eq:governing2DRescal}
\begin{align}
  \dotxp{\phi} &= \omega -\kappa\sin\phi \label{eq:governing2DRescalPhi},\\
  \dotxp{\kappa} &= \epsilon (1+a\cos\phi  - \kappa)\label{eq:governing2DRescalKap}.\
\end{align}
\end{subequations}

\subsubsection{Stability analysis}
Equilibrium conditions are given by Eqs.~\eqref{eq:governing2DRescalKap} and~\eqref{eq:governing2DRescalPhi} with
\begin{align}
    \kappa&=1+a\cos\phi\label{eq:EQcond2Done}
\end{align}
and
\begin{align}
    \omega = \kappa\sin{\phi}=(1 +  a  \cos \phi )  \sin \phi  \label{eq:EQcond2Dtwo}.\
\end{align}
Using Euler's identity, we may eliminate $\phi$ to obtain an equilibrium condition in $\kappa$ only,
\begin{align}\label{eq:EQcond2Dkappa}
    1 &= \frac{\omega^2}{\kappa^2} + \frac{(\kappa-1)^2}{a^2}.
\end{align}
The equilibria for $\phi$ are given as
\begin{align}\label{eq:EQcond2Dphi}
    \tan{\phi} &= \frac{\omega}{\kappa}\frac{a}{\kappa-1}
\end{align}
as a function of $\kappa$. Equilibria (and their bifurcations) are shown in Fig.~\ref{fig:2Dbif} for varying values of $\omega$ and $a$.

To determine the asymptotic stability of these equilibria, we consider the Jacobian of~\eqref{eq:governing2DRescal},
\begin{align}
  \label{eq:Jac2D}
  J &= - \begin{bmatrix}
    \kappa\cos\phi & \sin\phi\\
    \epsilon a\sin\phi & \epsilon
  \end{bmatrix}.
  \
\end{align}
Using equilibrium conditions \eqref{eq:EQcond2Done} and \eqref{eq:EQcond2Dtwo},
we can eliminate $\phi$ to obtain the trace and determinant for equilibria,
\begin{align}
  \det{(J^*)} &=  \epsilon(\kappa(\kappa-1)/a - a\omega^2/\kappa^2),\label{eq:detJEQ}\\
  \tr{(J^*)} &= \kappa(1-\kappa)/a - \epsilon.\
  \label{eq:trJEQ}
\end{align}

We seek to determine bifurcation curves in $(\omega,a)$-parameter-space.
Our system is planar so that saddle-node and Hopf bifurcations are determined via the conditions $\det{(J)}=0$ and $\tr{(J)}=0$ with $\det{(J)}>0$, respectively. Bogdanov-Takens bifurcations occur at the intersection of Hopf and saddle-node bifurcations.
Eliminating $\phi$ and $\kappa$ corresponding to equilibria in $\det{(J)}$ or $\tr{(J)}$ appears not to be feasible since \eqref{eq:EQcond2Dkappa} neither can be solved for $\kappa$ in closed form nor is there a suitable substitution to achieve the elimination. Instead, to determine the desired bifurcation curves we seek a parameterization of $\omega$ and $a$ in terms of one of the equilibrium variables, $\kappa$.

\paragraph{Saddle-node bifurcation curve.}
To determine such a parameterization, instead of using the condition $\det{J}=0$ for saddle-node bifurcations, we pursue another strategy.
Solving the equilibrium condition \eqref{eq:EQcond2Dkappa} for
\begin{align}
    a(\kappa) &=\pm  \frac{\kappa(\kappa-1)}{\sqrt{\kappa^2-\omega^2}}\label{eq:aAtEQ2D}\
\end{align}
and computing
\begin{align}
    \frac{\txtd a}{\txtd\kappa} &= \pm(\kappa^3+\omega^2(1-2\kappa))(\kappa^2-\omega^2)^{2/3},\
\end{align}
we find that $\txtd a/\txtd\kappa=0$ corresponds to a saddle-node condition,  resulting in
\begin{align}
\omega_{\text{SN}}(\kappa) = \pm\frac{\kappa^{3/2}}{\sqrt{2\kappa-1}}.\label{eq:omegaAtSN2D}
\end{align}

With \eqref{eq:omegaAtSN2D} and \eqref{eq:aAtEQ2D} (where \eqref{eq:omegaAtSN2D} is inserted for $\omega$), we have obtained the desired parameterization $(\omega_{\text{SN}}(\kappa),a_{\text{SN}}(\kappa))$ for the saddle-node bifurcation.

\paragraph{Hopf bifurcation curve.}
Solving
$\tr{(J^*)}=0$ for $a$ in \eqref{eq:trJEQ}, we find that
\begin{align}\label{eq:aH}
a_\text{H}(\kappa) = \kappa(1-\kappa)/\epsilon.
\end{align}
From \eqref{eq:EQcond2Dkappa} we obtain the equilibrium value for
\begin{align}
  \omega_\text{H}(\kappa) = \frac{\kappa}{a}\sqrt{a^2-\kappa^2+2\kappa-1}\label{eq:omegaAtEQ2D}
\end{align}
Hopf curves are thus determined via \eqref{eq:aH} and \eqref{eq:omegaAtEQ2D} with $\kappa$ as parameter, provided that $\det{(J^*)}$ in~\eqref{eq:detJEQ} is positive.
Note that all Hopf bifurcations in this system are subcritical so the limit cycles created in the Hopf bifurcations are unstable.

\paragraph{Homoclinic and heteroclinic bifurcations.} When a limit cycle collides with a saddle point it gets destroyed in a homoclinic bifurcation. Homoclinic bifurcations have been determined numerically as follows. Since limit cycles in this planar system are unstable, Eqs.~\eqref{eq:governing2DRescal} have been numerically integrated backwards in time while continuing $\omega$. The associated homoclinic bifurcation has been determined for a given value of $a$ by determining the $\omega$ at which the limit cycle is destroyed in saddle collision. Doing so for several values of $a$ allowed to construct the associated homoclinic bifurcation curve shown in Fig.~\ref{fig:2Dstab}. The same applies to heteroclinic bifurcations.

\paragraph{Bogdanov-Takens points.}
Curves of saddle-node,  Hopf, and homoclinic bifurcations intersect in a codimension 2 Bogdanov-Takens bifurcation point (BT). This point is found by seeking solutions that simultaneously satisfy equilibrium conditions together with $\det{(J)}=\tr{(J)}=0$.

\paragraph{Cusp bifurcation points.}
Another codimension 2 bifurcation point is the cusp bifurcation (CP) where two saddle-node bifurcations meet.
Thus, we find that CP points are located at $(\omega_\text{CP},a_\text{CP})=(0,\pm 1)$.

\subsubsection{Stability and bifurcation diagrams}
\begin{figure*}
\begin{overpic}[width=.25\linewidth]{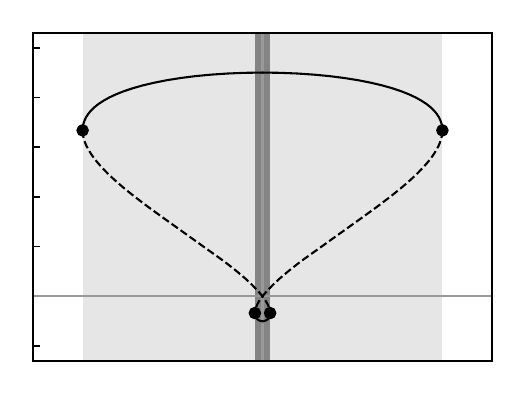}
  \lbltickTwoDBif
  \pnlblquarter{a}
  \pnhdln{$a=1.25$}
  \lblSN{20}{47}
  \lblSN{70}{47}
  \lblSN{35}{12}
  \lblSN{57}{12}
\end{overpic}
\nolinebreak
\begin{overpic}[width=.25\linewidth]{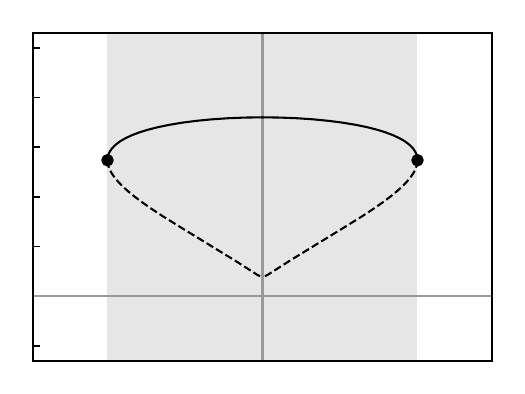}
  \lbltickTwoDBif
  \pnlblquarter{b}
  \pnhdln{$a=0.8$}
  \lblSN{25}{42}
  \lblSN{65}{42}
  \put(33,55){stable nodes}
  \put(40,30){saddles}
\end{overpic}
\nolinebreak
\begin{overpic}[width=.25\linewidth]{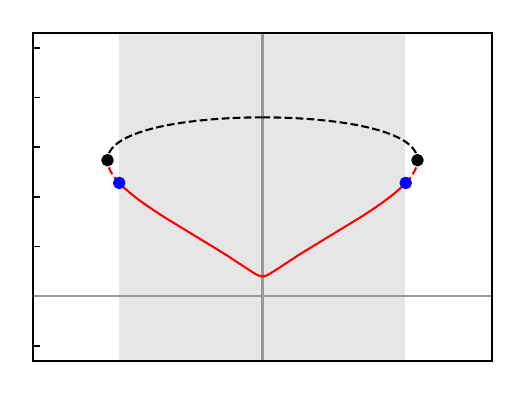}
  \lbltickTwoDBif
  \pnlblquarter{c}
  \pnhdln{$a=-0.8$}
  \lblSN{25}{42}
  \lblSN{65}{42}
  \lblHB{12}{32}
  \lblHB{77}{32}
  \put(40,55){saddles}
  \put(33,12){\textcolor{red}{stable spirals}}
\end{overpic} \nolinebreak
\begin{overpic}[width=.25\linewidth]{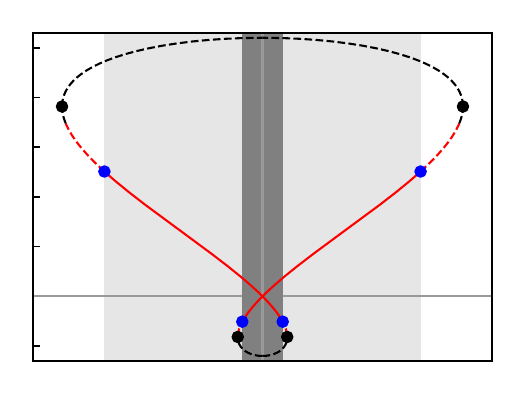}
  \lbltickTwoDBif
  \pnlblquarter{d}
  \pnhdln{$a=-1.6$}
  \lblSN{15}{53}
  \lblSN{76}{53}
  \lblHB{10}{35}
  \lblHB{82}{35}
  \lblSN{34}{8}
  \lblSN{57}{8}
  \lblHB{35}{15}
  \lblHB{57}{15}
\end{overpic}
\\[3mm]
\begin{overpic}[width=.25\linewidth]{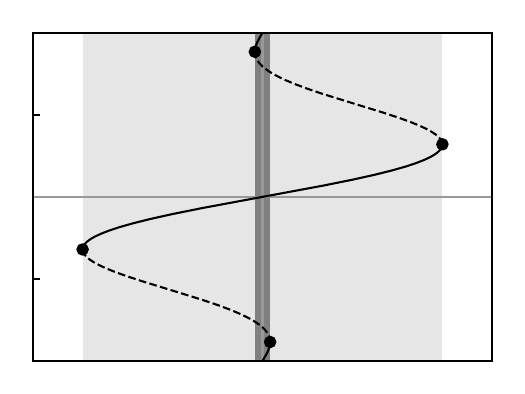}
  \lbltickTwoDBifPhi
  \put(24,40){In-phase}
  \put(10,63){Anti-phase}
  \put(55,10){Anti-phase}
\end{overpic}
\nolinebreak
\begin{overpic}[width=.25\linewidth]{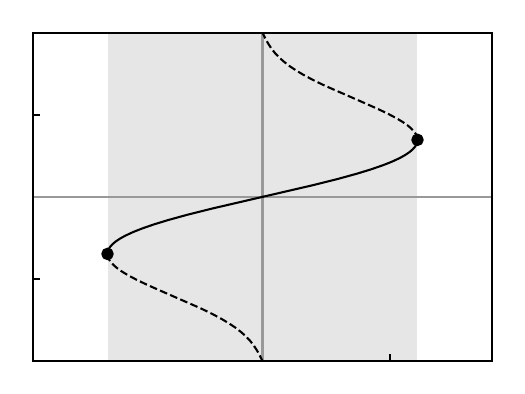}
  \lbltickTwoDBifPhi
\end{overpic}
\nolinebreak
\begin{overpic}[width=.25\linewidth]{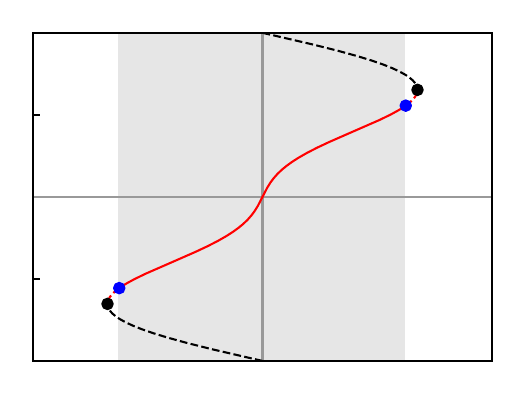}
  \lbltickTwoDBifPhi
\end{overpic}
\nolinebreak
\begin{overpic}[width=.25\linewidth]{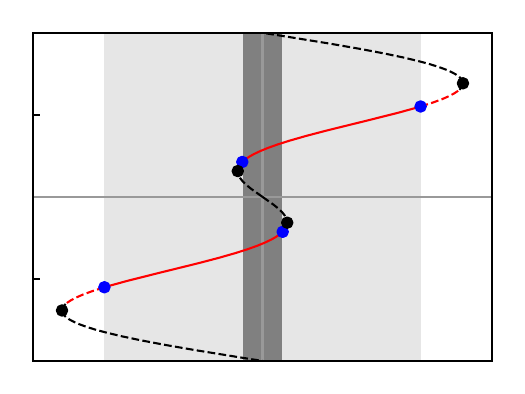}
  \lbltickTwoDBifPhi
\end{overpic}
\caption{Bifurcation diagrams (equilibria only) in $(\omega,\kappa)$ and $(\omega,\phi)$ for varying values of $a$ (indicated as \dIndicLine s in Fig.~\ref{fig:2Dstab}) with $\alpha=\beta=0$ and $\epsilon=0.2$. The diagrams show
stable nodes (\bdSTABLENODE s), saddles, unstable nodes (both \bdSADDLE s)
stable spirals (\bdSTABLEFOCUS s), and unstable spirals (\bdUNSTABLEFOCUS s).
Unstable limit cycles (librations; not shown) emerge from subcritical Hopf bifurcations. 
\highlight{SN and HB denote saddle-node and Hopf points, respectively.} }
\label{fig:2Dbif}
\end{figure*}
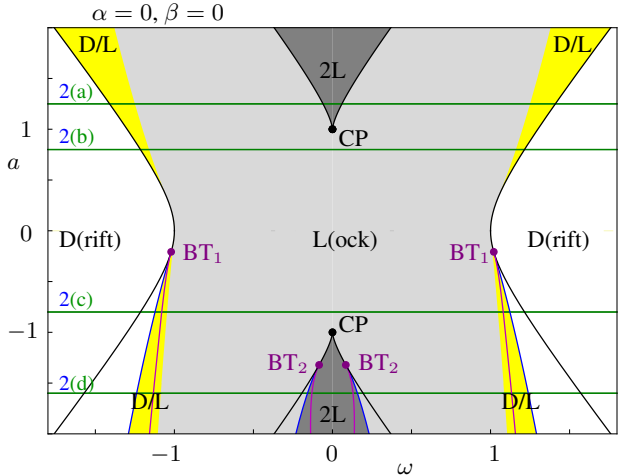
\begin{figure}
\begin{overpic}[width=\linewidth]{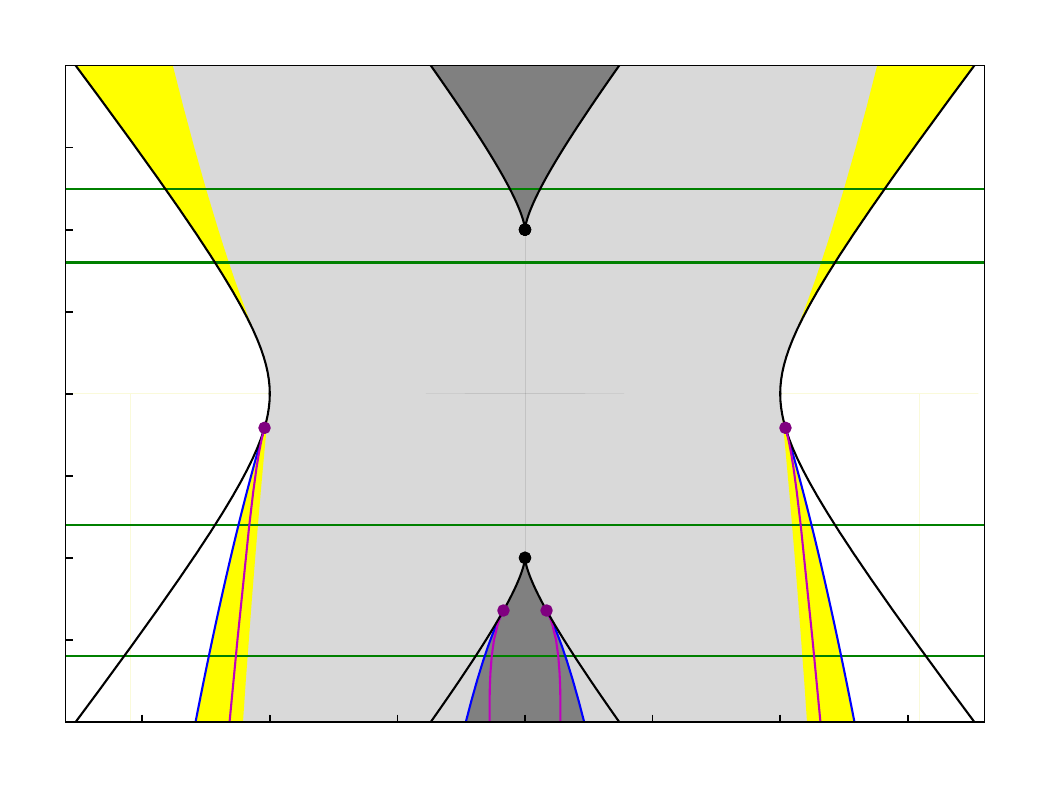}
\pnhdln{$\alpha=0$, $\beta=0$}
  \axlblhalf{$\omega$}{$a$}
  \put(2,52){$1$}
  \put(2,36){$0$}
  \put(0,21){$-1$}
  \put(22,2){$-1$}
  \put(49,2){$0$}
  \put(74,2){$1$}
  \put(8,35){\lbldr (rift)}
  \put(80,35){\lbldr (rift)}
  \put(48,61){\lblLL}
  \put(48,8){\lblLL}
  \put(47,35){\lblL(ock)}
  \put(11,65){\lblLdr}
  \put(84,65){\lblLdr}
  \put(19,10){\lblLdr}
  \put(76,10){\lblLdr}
  \lblCP{51}{50.5}
  \lblCP{51}{22}
  \lblBTone{27}{33}
  \lblBTone{68}{33}
  \lblBTtwo{40}{16}
  \lblBTtwo{54}{16}
  \pointertootherfig{8}{58}{\ref{fig:2Dbif}(a)}
  \pointertootherfig{8}{51}{\ref{fig:2Dbif}(b)}
  \pointertootherfig{8}{26}{\ref{fig:2Dbif}(c)}
  \pointertootherfig{8}{13}{\ref{fig:2Dbif}(d)}
\end{overpic}
\caption{Stability diagram in $(\omega,a)$ for $\alpha=\beta=0$ with $\epsilon=0.2$. $(\omega,a)$ parameter regions are shaded according to which stable states they contain: Drift (\lbldr) and/or locked states (\lblL). \highlight{BT and CP denote Bogdanov-Takens and Cusp co-dimension 2 bifurcation points.}}
\label{fig:2Dstab}
\end{figure}

We explain the bifurcation scenarios and associated stability boundaries based on our previous analysis.
We first observe that the bifurcation and stability diagram (see Figs.~\ref{fig:2Dbif} and~\ref{fig:2Dstab}, respectively) exhibit certain symmetries regarding the reflection $\omega\mapsto -\omega$.
It is easy to check that Eqs.~\eqref{eq:governing2DRescal}, ~\eqref{eq:EQcond2Dkappa} and \eqref{eq:EQcond2Dphi} obey the symmetry \eqref{eq:swapnames}
which effectively swaps the labels of the two oscillators and thus preserves equilibria, their stabilities and bifurcations (as seen in Figs.~\ref{fig:2Dbif} and \ref{fig:2Dstab}).
A further symmetry regarding $a \mapsto -a$ \eqref{eq:minusaphiinvshift} preserves equilibria, 
but stability of the equilibria is not preserved as can be seen by inspecting the Jacobian in \eqref{eq:Jac2D} (more specifically, saddle-node and cusp bifurcations are preserved, while Hopf bifurcations are not, see Figs.~\ref{fig:2Dstab} and~\ref{fig:2Dbif}(b),(c)). See also Appendix~\ref{app:symm} for further details and explanations.

The non-adaptive case ($ a = 0 $) represents the limit of the classical Kuramoto model (see Sec.~\ref{sec:Dyn1D}). This case illustrates the perhaps most fundamental bifurcation organizing the structure of the stability diagram on a 'global' scale. When intrinsic frequencies are sufficiently similar, $ \omega = \omega_1 - \omega_2 < 1$, oscillators lock their frequency, i.e., synchronization is achieved; vice versa, if intrinsic frequencies are too dissimilar, $ \omega > 1 $, oscillator phases drift apart. Close to $\omega=0$, oscillators are nearly in-phase (compare with Fig.~\ref{fig:2Dbif} where $a\neq 0$); but the phase difference $\phi$ increases with larger $|\omega|$ until the locking breaks apart in a saddle-node bifurcation (\sdSN) and a drifting state appears.
This saddle-node bifurcation separates the stability regions for frequency-locked (\lblL, \sdL) and drifting (\lbldr, \sddr) solutions and can be followed throughout the stability diagram in Fig.~\ref{fig:2Dstab}. Note that this drift solution corresponds to a rotation on the cylinder $\T\times \R$.

For non-zero adaptivity ($ a\neq 0 $), the structure of the stability diagram in Fig.~\ref{fig:2Dstab} is organized around two bifurcation points of codimension two, from where additional bifurcation curves emanate:
i) cusp points (CP, \sdCP) located at $(\omega_\text{CP},a_\text{CP})=(0,\pm 1)$; and
ii) Bogdanov-Takens points (BT, \sdBT).
Furthermore, several parameter regions of bistability (\sdLL s, \sdLdr s) appear near the cusp bifurcation points (\sdLL s) and the saddle-node bifurcations leading to drifting solutions (\sdLdr s).

\paragraph{Positive adaptivity ($a>0$). }
For smaller frequency mismatch (around $\omega <1$)
and intermediate adaptivity ($0<a<1$), frequency-locked solutions (\lblL, \sdL in Fig.~\ref{fig:2Dstab}) with positive coupling strengths exist, stable nodes (\bdSTABLENODE\ in Fig.~\ref{fig:2Dbif} (b)) closer to in-phase and saddles (\bdSADDLE) closer to anti-phase configurations.
For $a\geq 1$, two additional saddle-node bifurcation curves (\bdSN s in Fig.~\ref{fig:2Dbif} (a) / \sdSN s in Fig.~\ref{fig:2Dstab}) emerge in a cusp bifurcation (CP, \sdCP\ in Fig.~\ref{fig:2Dstab}) located at $(\omega_\text{CP},a_\text{CP})=(0,1)$. Traversing these saddle-node bifurcation curves gives birth to an additional stable node and a saddle with negative coupling, see Fig.~\ref{fig:2Dbif} (a). Thus, for $a>1$,  above the cusp point, between the inner SN bifurcation curves, we have a region of bistability (\sdLL) with two stable locked solutions characterized by near in- and anti-phase configurations, respectively.

For larger frequency mismatch (around $ \omega > 1 $), as we traverse the saddle-node bifurcation giving rise to the lock state (\lblL), we first observe a region of bistability (\sdLdr) between the lock state (\lblL) and a drift cycle (\lbldr).
As the frequency mismatch $|\omega|$ is further diminished, the drift cycle collides with the saddle version of the lock state emanating from the saddle-node bifurcation and is thus destroyed in a heteroclinic bifurcation.

\paragraph{Negative adaptivity ($ a < 0 $). }
The symmetry~\eqref{eq:minusaphiinvshift} swaps near in-phase with anti-phase configurations and changes the stability of equilibria. The overall bifurcation structure is thus similar to the one observed for $ a > 0 $; albeit, their structure is more intricate.

We first restrict our attention to smaller frequency mismatch ($ \omega < 1 $).
For $-1<a<0$, the saddle-node bifurcations (\bdSN\ in Fig.~\ref{fig:2Dbif} (c) / \sdSN\ in Fig.~\ref{fig:2Dstab}) now give birth to a saddle and an unstable node. The unstable node turns into an unstable spiral which later gains stability in a Hopf bifurcation (HB). The resulting stable spiral (\bdSTABLEFOCUS\ in Fig.~\ref{fig:2Dbif} (c)) now, however,  corresponds to a near in-phase state with positive coupling (but smaller compared to $a>0$).
We observe a cusp bifurcation point (CP, \sdCP\ in Fig.~\ref{fig:2Dstab}) at $(\omega_\text{CP},a_\text{CP})=(0,-1)$. Below the cusp point ($a<-1$), a Bogdanov-Takens point (BT$_2$, \sdBT\ in Fig.~\ref{fig:2Dstab}) appears on the associated saddle-node bifurcation (\bdSN\  in Fig.~\ref{fig:2Dbif} (d) / \sdSN\ in Fig.~\ref{fig:2Dstab}) which gives rise to a subcritical Hopf bifurcation (\bdHB\ in Fig.~\ref{fig:2Dbif} (d) / \sdHB\ in Fig.~\ref{fig:2Dstab}) that stabilizes the second lock state, a stable spiral with small negative coupling and $\phi$ near $\pm \pi/4$ (\bdSTABLEFOCUS\ in Fig.~\ref{fig:2Dbif} (d)). The unstable limit cycles (librations) emanating from the Hopf bifurcation get destroyed in a homoclinic bifurcation (\sdHC).

For larger frequency mismatch ($ \omega > 1 $), we observe a region of bistability (\lblLdr, \sdLdr\ in Fig.~\ref{fig:2Dstab}) between lock (\lblL) and drift (\lbldr) states  similar to the one observed for positive adaptivity, $a>0$. However, there are important differences.
First, below the Bogdanov-Takens point BT$_1$, the frequency-locked solution (\lblL) loses stability in a subcritical Hopf bifurcation (\sdHB\ in Fig.~\ref{fig:2Dstab}).
More specifically, as shown in Fig.~\ref{fig:2Dbif}(c),(d), the saddle-node bifurcation (\bdSN) gives birth to a saddle (\bdSADDLE) and an unstable node which first becomes an unstable spiral (\bdUNSTABLEFOCUS) and then becomes a stable spiral (\bdSTABLEFOCUS) in a Hopf bifurcation (\bdHB).

Second, for adaptivity values roughly above the BT$_1$ point, 
the drift cycle \lbldr\ is destroyed in a heteroclinic bifurcation;
this scenario, however, appears to be quite different for adaptivity $a$ below BT$_1$.
On the boundary of the \lblLdr\ (\sdLdr) and \lblL\ regions (\sdL) --- before any  heteroclinic bifurcation is possible to occur  --- the drift (limit) cycle loses its stability. Thus, trajectories emanating from the unstable drift end up spiraling into the stable spiral that arises in the Hopf bifurcation related to BT$_1$.

To summarize the respective differences regarding stability changes occurring for $a<0$ and $a>0$, the presence of Bogdanov-Takens bifurcation points BT$_1$ and BT$_2$ diminishes the size of the \lblL\ locking region and the \lblLL\ bistability region (\sdLL\ in Fig.~\ref{fig:2Dstab}). Furthermore, the size of the bistable  \lblLdr\ region with lock and drift is also effectively diminished.

\subsection{\highlight{Asymmetric adaptation}}\label{sec:Dyn3D}
We now allow for arbitrary values of $\beta$; in general, this may lead to asymmetric coupling strengths, $\kappa_{12}\neq\kappa_{21}$, corresponding to directed network weights.

\subsubsection{Stability analysis}

We first investigate equilibria of the full three dimensional system in Eqs.~\eqref{eq:governingRescal} which gives rise to
the following fixed point conditions:
\begin{subequations}\label{eq:EQcond3D}
\begin{align}
2\omega &=  \kappa_{21}\sin(\alpha+\phi) - \kappa_{12}\sin(\alpha-\phi),\\
\kappa_{12}&=1+a\cos(\beta+\phi)\label{eq:EQcond3DKap12}\\
\kappa_{21}&=1+a\cos(\beta-\phi)\label{eq:EQcond3DKap21}
\end{align}
\end{subequations}
Eliminating $\kappa_{12}$ and $\kappa_{21}$, we obtain a condition that only depends on $\phi$ and guarantees the existence of an equilibrium,
\begin{align}
\omega = (\cos \alpha +  a  \cos (\alpha -\beta )\cos \phi )  \sin \phi.  \label{eq:EQcond3DPhi}
\end{align}
Note that Eqs.~\eqref{eq:EQcond3DKap12}, \eqref{eq:EQcond3DKap21} and \eqref{eq:EQcond3DPhi} provide us
with a parameterization of an equilibrium curve for $(\omega,\kappa_{12},\kappa_{21})$ in variable $\phi$.

To determine the stability of equilibria, we calculate the Jacobian for~\eqref{eq:governingRescal},
\begin{align}\label{eq:Jac3D}
  J&=\begin{bmatrix}
  J_{\phi\phi} &
  \frac{1}{2}\sin(\alpha-\phi) &
  -\frac{1}{2}\sin(\alpha+\phi) \\
  -\epsilon a\sin(\beta+\phi) &
  -\epsilon &
  0 \\
  \epsilon a\sin(\beta-\phi) &
  0 &
  -\epsilon
  \end{bmatrix},\notag\
\end{align}
where $J_{\phi\phi} = -(\kappa_{12}\cos(\phi-\alpha)+\kappa_{21}\cos(\phi+\alpha))/2$.
The Jacobian has the eigenvalues 
\begin{subequations}\label{eq:Jac3Deval}
  \begin{align}
  \lambda_{1,2}&=\mu\pm\sqrt{\delta},\\
  \lambda_3&=-\epsilon,\
  \end{align}
\end{subequations}
where
\begin{subequations}\label{eq:Jac3Dexpr}
\begin{align}
    \mu &= -\frac{A}{4} -\frac{\epsilon}{2},\label{eq:Jac3Dmu}\\
    \delta &= \frac{\epsilon}{2}(a B - A) +\frac{1}{16}(A+2 \epsilon)^2,\label{eq:Jac3Ddelta}\\
    A&:=\kappa_{12}\cos (\alpha -\phi )+\kappa_{21}\cos (\alpha +\phi ),\label{eq:Jac3D_A}\\
    B&:=\cos (\alpha +\beta )-\cos{ 2 \phi}  \cos {(\alpha -\beta )}.\label{eq:Jac3D_B}\
\end{align}
\end{subequations}
Since by definition $\lambda_3<0$, it suffices to only consider $\lambda_{1,2}$.

If an equilibrium point satisfies $\mu^2=\delta$, one eigenvalue becomes zero, and the equilibrium is a saddle-node point. One could use this condition to determine the associated saddle-node curves, but we instead use the following consideration.
Since the equilibrium condition \eqref{eq:EQcond3DPhi} only depends on $\phi$, the conditions for an equilibrium and for a saddle-node bifurcation are reduced to a problem in a single variable, $\phi$. Accordingly, we require in addition to \eqref{eq:EQcond3DPhi} that $\txtd \omega / \txtd \phi = 0$. Solving the resulting two conditions for $(\omega,a)$ results in a parameterization of the saddle-node curves in $\phi$,
\begin{subequations}
  \begin{align}
  \omega_{\text{SN}} &= -\cos\alpha\sec2\phi\sin^3\phi,\\
  a_{\text{SN}} &= -\cos\alpha\cos\phi\sec(\alpha-\beta)\sec2\phi.
  \end{align}
\end{subequations}

Next, we consider equilibria undergoing Hopf bifurcations. If a given equilibrium satisfies $\mu=0,\delta<0$, the real part of a complex conjugated pair of eigenvalues vanishes and the equilibrium is a Hopf point.
We find a parameterization for the Hopf curves as follows. First, we require that the equilibrium condition in \eqref{eq:EQcond3DPhi} is satisfied. Second, after substituting \eqref{eq:EQcond3DKap12} and \eqref{eq:EQcond3DKap21} into \eqref{eq:Jac3D_A} we require that $\mu=0$. Solving the two resulting conditions for $(\omega,a)$ we obtain
\begin{subequations}
\begin{align}
\omega_{\text{H}} &=
\frac{
  (\epsilon  \cos \alpha  \cos \beta  \cos \phi + (\cos \alpha +\epsilon  \cos \phi )\sin \alpha  \sin \beta )\sin \phi
}
{\sin\alpha\sin\beta\sin^2\phi-\cos\alpha\cos\beta\cos^2\phi},
\\
a_{\text{H}} &= \frac{
\epsilon + \cos\alpha\cos\phi
}{
\sin\alpha\sin\beta\sin^2\phi-\cos\alpha\cos\beta\cos^2\phi
}.\
\end{align}
\end{subequations}
Now $(\omega_{\text{H}},a_{\text{H}})$ with parameter $\phi$ delineates a Hopf curve provided that $\delta<0$ in \eqref{eq:Jac3Ddelta}.

Finally, if $\mu=\delta=0$, two eigenvalues are zero and the equilibrium is a Bogdanov-Takens  point.
Thus, simultaneously solving $\mu=\delta=0$ together with the fixed point conditions \eqref{eq:EQcond3D},  given a value of $\epsilon$ we are able to obtain $(\phi,\kappa_{12},\kappa_{21},a,\omega)$ at a Bogdanov-Takens point.

\subsubsection{Dynamics for $\beta=\pi/2$}\label{sec:beta0_5pi}

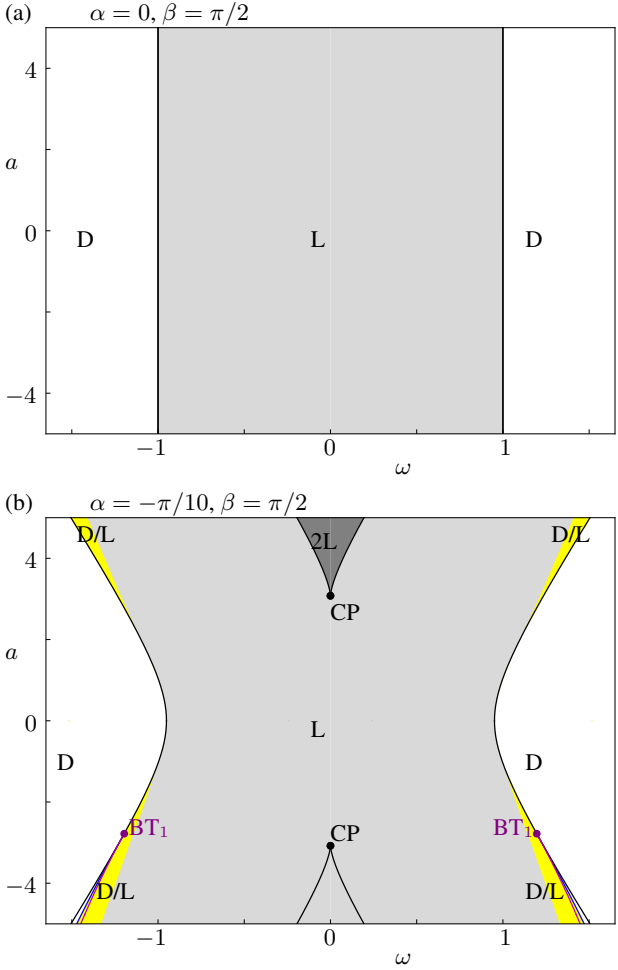
\begin{figure}
\begin{overpic}[width=\linewidth]{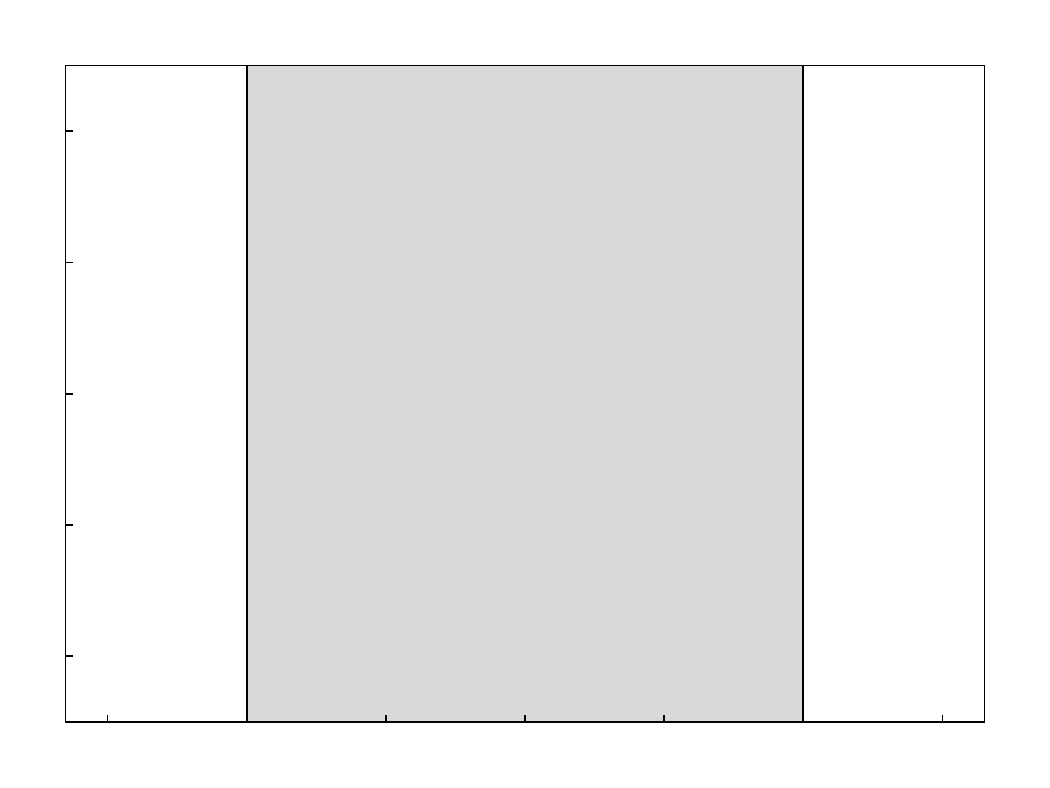}
\pnhdln{$\alpha=0$, $\beta=\pi/2$}
\axlblhalf{$\omega$}{$a$}
\pnlblhalf{a}
\put(3,61){$4$}
\put(3,36){$0$}
\put(0,11){$-4$}
\put(49,3){$0$}
\put(20,3){$-1$}
\put(76,3){$1$}
\put(47,35){\lblL}
\put(11,35){\lbldr}
\put(80,35){\lbldr}
\end{overpic}
\begin{overpic}[width=\linewidth]{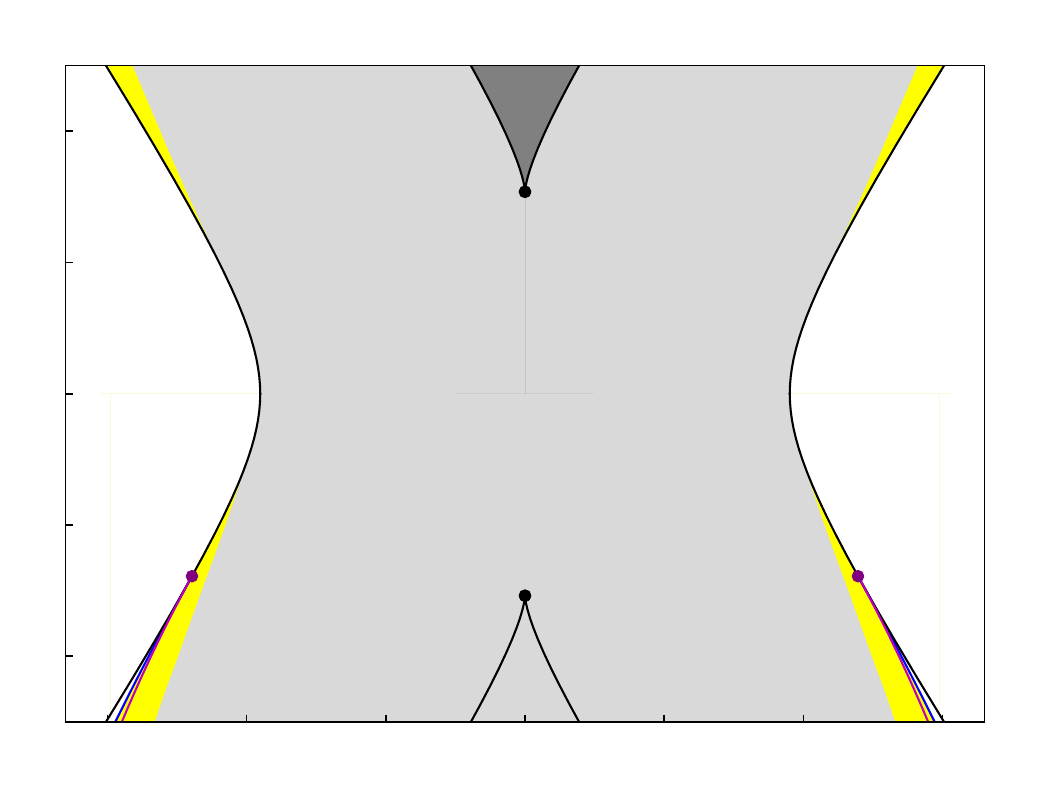}
\pnhdln{$\alpha=-\pi/10$, $\beta=\pi/2$}
\pnlblhalf{b}
\axlblhalf{$\omega$}{$a$}
\put(3,61){$4$}
\put(3,36){$0$}
\put(0,11){$-4$}
\put(49,3){$0$}
\put(20,3){$-1$}
\put(76,3){$1$}
\put(8,30){\lbldr}
\put(80,30){\lbldr}
\put(47,64){\lblLL}
\put(47,35){\lblL}
\put(11,65){\lblLdr}
\put(84,65){\lblLdr}
\put(14,10){\lblLdr}
\put(80,10){\lblLdr}
\lblCP{50}{53}
\lblCP{50}{19}
\lblBTone{19}{20}
\lblBTone{75}{20}
\end{overpic}
\caption{
Stability diagrams for $\beta=\pi/2$.
$(\omega,a)$ parameter regions are shaded according to which stable states they contain: Drift (\lbldr) and/or locked states (\lblL).
}
\label{fig:stabbeta0_5pi_alpham0_1pi}
\end{figure}
\highlight{
Substituting $\beta=\pi/2$ into Eqs.~\eqref{eq:sumdiff}, we have 
\begin{subequations}
    \begin{align}\label{eq:pihalfDeltaSigma}
        \dotxp{\Delta} &= -\epsilon(2a\sin(\phi)+\Delta),\\
        \dotxp{\Sigma} &= \epsilon(2-\Sigma),
    \end{align}
\end{subequations}
so that $\Sigma \to 2$ as $t\to\infty$ and the system is attracted to a two dimensional subspace on which the dynamics are given by
\begin{subequations}
  \begin{align}
    \dotxp{\phi} &= \omega + \frac{\sin{\alpha}}{2} \Delta \cos{\phi} - \cos{\alpha}\sin{\phi},\\
    \dotxp{\Delta} &= -\epsilon(2a\sin(\phi)+\Delta).
  \end{align}
\end{subequations}
Note that in the asymptotic time limit $t\to\infty$ the coupling is not anti-symmetric, i.e., $\kappa_{12}=\kappa_{21}$; this would be the case if instead we had chosen $a_0 = 0$ in our model Eqs.~\eqref{eq:adaptationrule}. Instead, we have $\kappa_{12}+\kappa_{21} \to 2$.
}

\paragraph{Phase-lag $\alpha=0$ and $\alpha=\pi$.} 
\highlight{
We first discuss the simplest case with $\alpha=0$. 
The asymptotic dynamics in 
\begin{subequations}
  \begin{align}
    \dotxp{\phi} &= \omega - \sin\phi,\\
    \dotxp{\Delta} &= -\epsilon(2a\sin\phi+\Delta).\
  \end{align}
\end{subequations}
While the variable $\phi$ drives $\Delta$, its dynamics is independent of $\Delta$. Thus, we effectively observe the one-dimensional dynamics known for the Kuramoto model
\eqref{eq:governing1DRescal} discussed in Sec.~\ref{sec:Dyn1D}.
}

\highlight{For $\omega < 1$ we observe locked states with $\omega=\sin\phi$.
The resulting equilibria are
\begin{subequations}
\begin{align}
\phi&=\begin{cases}
        \pi-\arcsin\omega\\
        \arcsin\omega,\
      \end{cases}
      \\
\kappa_{12}&=1-a\omega,\\
\kappa_{21}&=1+a\omega.\
\end{align}
\end{subequations}
where the latter near in-phase state is stable.
The equilibrium condition, $\omega=\sin{\phi}$, informs us that a saddle-node bifurcation occurs for $ \omega = \pm 1$, regardless of the value of $a$. The associated \highlight{stability boundaries} are therefore the  straight \sdSN s shown in Fig.~\ref{fig:stabbeta0_5pi_alpham0_1pi}(a).
Thus, contrary to other parameter choices considered in this study, the dynamics and bifurcations do not increase in complexity as $a$ is varied: we are dealing with a special limiting case. The equilibria are the same for $\alpha=\pi$, however, with reversed stability.
}

\paragraph{Arbitrary phase-lag ($\alpha \neq 0,\pi$).} 
It is easily checked that \eqref{eq:pihalfDeltaSigma} has the parameter symmetry $(a,\alpha)\mapsto(-a,-\alpha)$. Since we were allowed to limit $\alpha\in(-\pi/2,\pi/2]$ in the original governing equations~\eqref{eq:governingRescal}, we may restrict $\alpha$ further to either  $[-\pi/2,0]$ or $[0,\pi/2]$ while observing this symmetry.

To keep the analysis manageable,
we consider only $\alpha=-\pi/10$. The resulting stability diagram is shown in Fig.~\ref{fig:stabbeta0_5pi_alpham0_1pi}(b).
Even though this is just a small deviation, the bifurcation landscape differs drastically from the case where $\alpha=0$.  As was the case for $\alpha=\beta=0$ (Fig.~\ref{fig:2Dstab}), the stability diagram is symmetric regarding $\omega\mapsto - \omega$ (preserving equilibria and their stabilities and bifurcations, see \eqref{eq:swapnames}). Moreover, equilibria, and the SN and cusp bifurcations are symmetric about $a\mapsto - a$ (preserving, equilibria, and the SN and cusp bifurcations, see \eqref{eq:minusaphiinvshift}). See Appendix~\ref{app:symm} for further details and explanations.

\subsubsection{Dynamics for $\beta=\pi/4$}\label{sec:beta0_25pi}
\begin{figure}
\begin{overpic}[width=\linewidth]{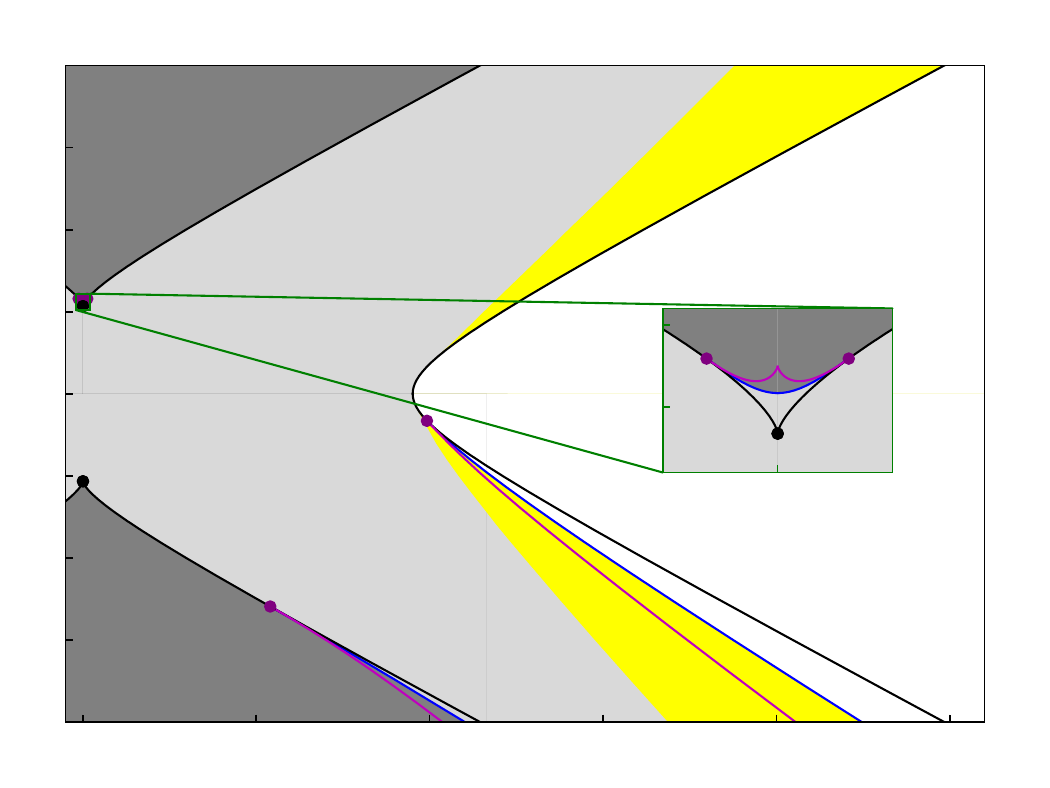}
\pnhdln{$\alpha=\pi/10$, $\beta=\pi/4$}
  \axlblhalf{$\omega$}{$a$}
  \put(3,60){$3$}
  \put(3,36){$0$}
  \put(0,13){$-3$}
  \put(8,3){$0$}
  \put(72,3){$2$}
  \insetaxtick{73}{28}{$0$    }
  \insetaxtick{82}{28}{$0.02$ }
  \insetaxtick{61}{28}{$-0.02$}
  \insetaxtick{59}{35}{$1.1$  }
  \insetaxtick{59}{43}{$1.2$  }
  \put(24,32){\lblL}
  \put(10,10){\lblLL}
  \put(10,60){\lblLL}
  \put(80,33){\lblL}
  \put(70,42){\lblLL}
  \put(80,20){\lbldr}
  \put(65,10){\lblLdr}
  \put(75,65){\lblLdr}
  \lblBTtwo{26}{18}
  \lblBTone{35}{32}
  \lblBTthree{63}{37}
  \lblBTthree{80}{37}
\end{overpic}
\caption{Stability diagram for $\alpha=\pi/10,\beta=\pi/4$. $(\omega,a)$ parameter regions are shaded according to which stable states they contain: Drift (\lbldr) and/or locked states (\lblL).}
\label{fig:stabbeta0_25pi_alpha0_1pi}
\end{figure}
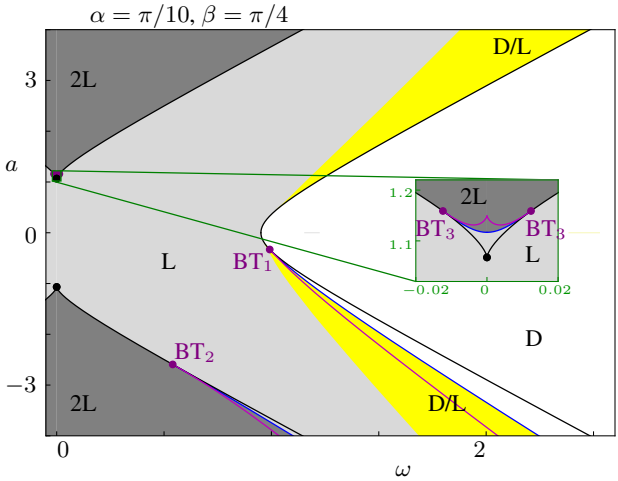
Next we consider the case where $\beta=\pi/4$. The stability diagram for $\alpha=0$ (not shown) is qualitatively identical to the one obtained for $\alpha = \beta = 0$ (Fig.~\ref{fig:2Dstab}). Therefore, we instead consider small deviations from $\alpha=0$, i.e., $\alpha=\pm\pi/10$.

\paragraph{Phase-lag $\alpha=\pi/10$.}

First, we consider
the stability diagram for $\alpha=\pi/10$, see Fig.~\ref{fig:stabbeta0_25pi_alpha0_1pi}.
There are three distinct Bogdanov-Takens points (BT$_1$ and BT$_2$ in the main plot of Fig.~\ref{fig:stabbeta0_25pi_alpha0_1pi} and (two) BT$_3$ in the inset) which organize the bifurcation structure.
The Hopf curves (\sdHB s) emanating from all three Bogdanov-Takens points (BT$_1$, BT$_2$, BT$_3$) are subcritical.
These Hopf curves are adjacent to SN curves (\sdSN s) giving birth to saddles and unstable nodes.
The unstable nodes turn into unstable spirals, and, when undergoing the Hopf bifurcations, into stable spirals. The homoclinic curves (\sdHC s) adjacent to the SN curves destroy the unstable (libration) limit cycles created in the subcritical Hopf bifurcations.

\paragraph{Phase-lag $\alpha=-\pi/10$.}
\begin{figure*}
\begin{overpic}[width=.5\linewidth]{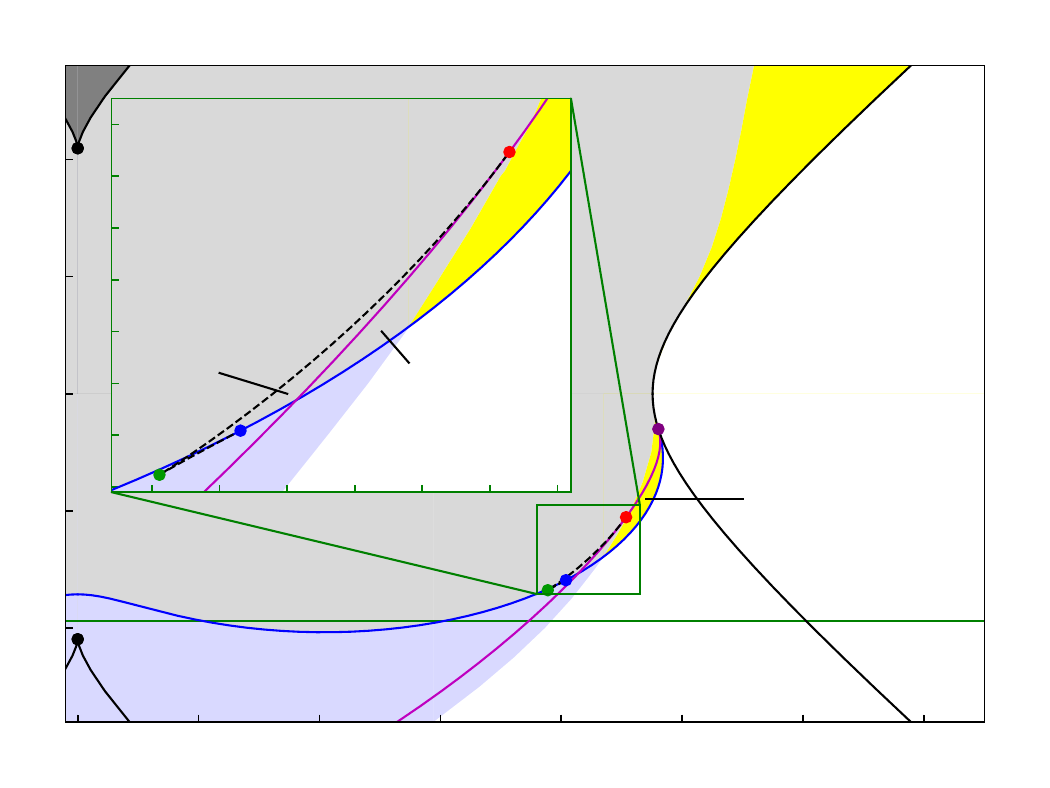}
\pnlblhalf{a}
\axlblhalf{$\omega$}{$a$}
\put(39,39){\lblLockLibrationAndMixedOscnoLC}
\put(13,40){\lblLockLibrationAndMixedOsc}
\put(20,10){\lblLibrationAndMixedOsc}
\put(20,30){\lblLibrationAndMixedOsc}
\put(20,55){\lblL}
\put(20,20){\lblL}
\put(6,66){\lblLL}
\put(71,26){\lblLdr}
\put(71,60){\lblLdr}
\put(50,60){\lblLdr}
\put(85,45){\lbldr}
\put(40,30){\lbldr}
\insetaxtick{19}{26}{$0.8$}
\insetaxtick{45}{26}{$0.9$}
\insetaxtick{7}{62}{$-1$}
\insetaxtick{4}{33}{$-1.6$}
\lblCP{7}{15}
\lblCP{7}{57}
\lblCPC{11}{31}
\lblGH{16}{35}
\lblSLH{40}{60}
\lblBTone{64}{33}
\put(3,58){$2$}
\put(3,36){$0$}
\put(0,13){$-2$}
\put(7,3){$0$}
\put(63,3){$1$}
\pointertootherfig{60}{14}{\ref{fig:alpham0_1pi_beta0_25pi_stabbif}(b)-(d),\,\,\ref{fig:alpham0_1pi_beta0_25pi_phsprt}(a)-(h)}
\end{overpic}
\nolinebreak
\begin{overpic}[width=.5\linewidth]{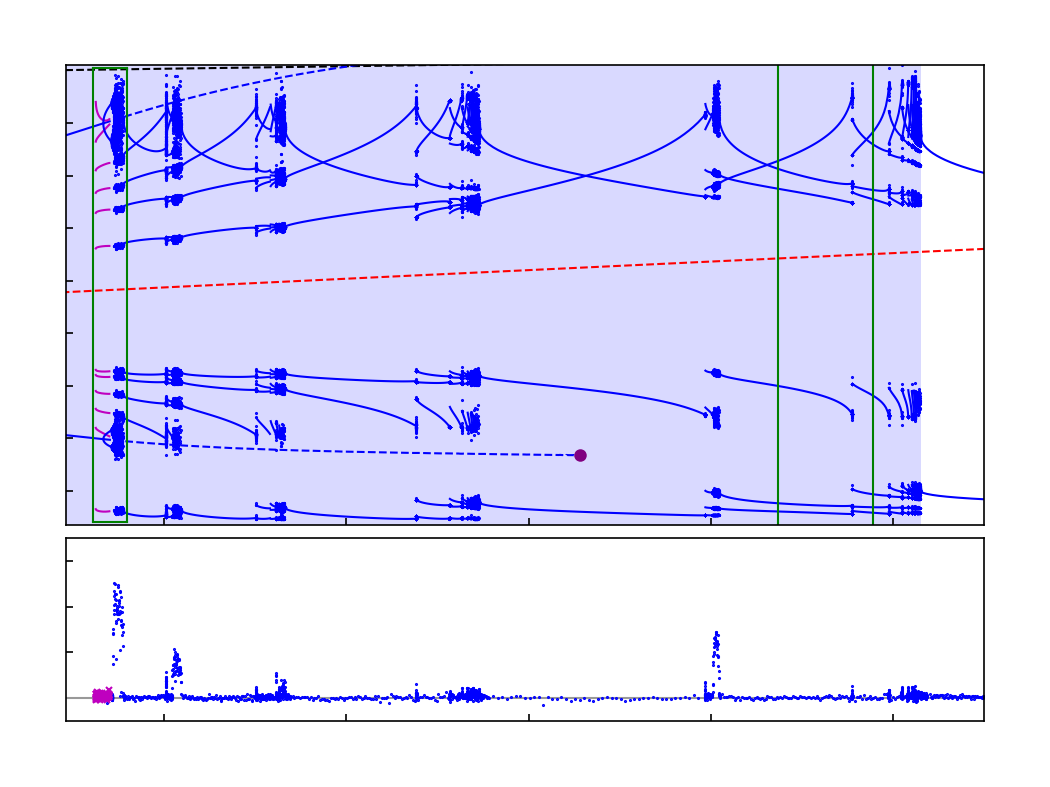}
\pnlblhalf{c}
\axlblhalf{$\omega$}{$\kappa_{12}$}
\put(50,70){\lblLibrationAndMixedOsc}
\put(90,70){\lbldr}
\pointertootherfig{12}{30}{(d)                                       }
\pointertootherfig{74}{30}{\ref{fig:alpham0_1pi_beta0_25pi_phsprt}(f)}
\pointertootherfig{84}{30}{\ref{fig:alpham0_1pi_beta0_25pi_phsprt}(g)}
\lblHC{57}{31}
\put(1,62){$2.6$}
\put(1,27){$1.2$}
\put(15,3){$0.7$}
\put(82,3){$0.78$}
\put(0,15) {$\maxLE$}
\put(4,7) {$0$}
\put(-1,20) {$0.03$}
\end{overpic}
\\[.5mm]
\begin{overpic}[width=.5\linewidth]{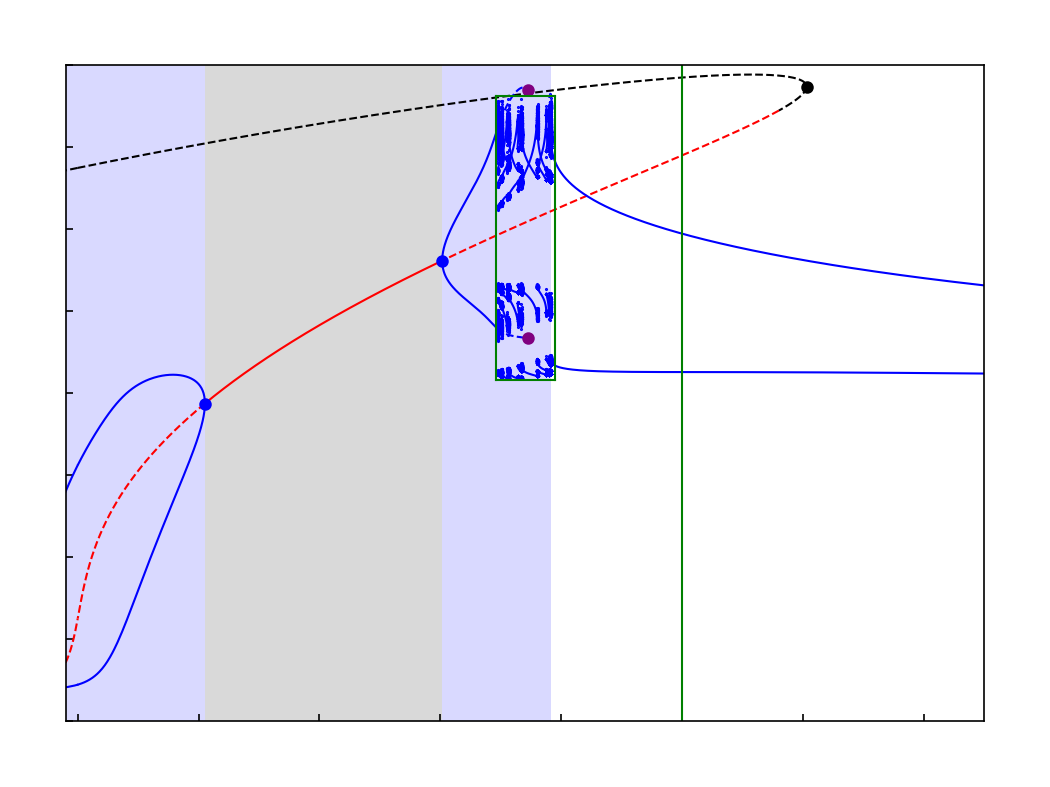}
\pnlblhalf{b}
\axlblhalf{$\omega$}{$\kappa_{12}$}
\put(7,3){$0$}
\put(63,3){$1$}
\put(3,21){$0$}
\put(3,37){$1$}
\put(3,53){$2$}
\put(10,70){\lbllibration}
\put(30,70){\lblL}
\put(42,70){\lblLibrationAndMixedOsc}
\put(60,70){\lbldr}
\lblHB{21}{35}
\lblHB{36}{50}
\lblHC{53}{42}
\lblHC{53}{65}
\lblSN{78}{65}
\pointertootherfig{53}{52}{(c)}
\pointertootherfig{65}{20}{\ref{fig:alpham0_1pi_beta0_25pi_phsprt}(h)}
\end{overpic}
\nolinebreak
\begin{overpic}[width=.5\linewidth]{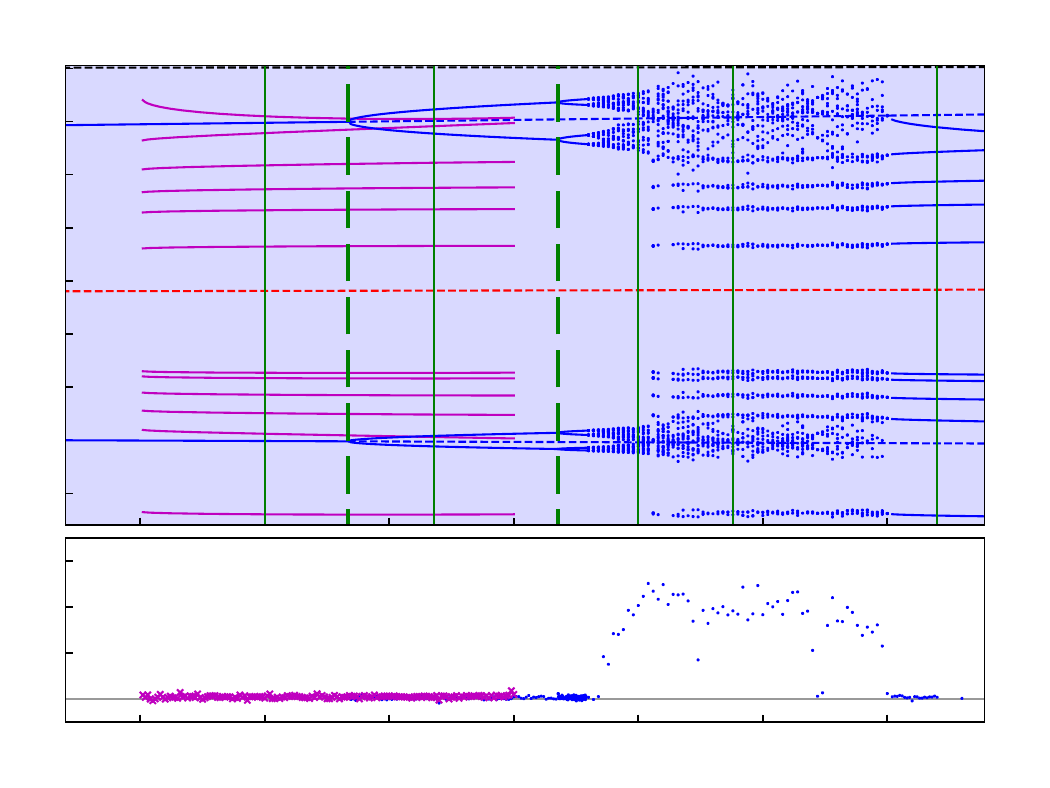}
\pnlblhalf{d}
\put(50,70){\lblLibrationAndMixedOsc}
\axlblhalf{$\omega$}{$\kappa_{12}$}
\put(1,62){$2.6$}
\put(1,27){$1.2$}
\put(21,3) {$0.693$}
\put(69,3) {$0.695$}
\put(28,29){\textcolor{dgreen}{PD}}
\put(48,29){\textcolor{dgreen}{PD}}
\put(0,15) {$\maxLE$}
\put(4,7) {$0$}
\put(-1,20) {$0.03$}
\pointertootherfig{20}{40}{\ref{fig:alpham0_1pi_beta0_25pi_phsprt}(a)}
\pointertootherfig{37}{40}{\ref{fig:alpham0_1pi_beta0_25pi_phsprt}(b)}
\pointertootherfig{57}{40}{\ref{fig:alpham0_1pi_beta0_25pi_phsprt}(c)}
\pointertootherfig{66}{40}{\ref{fig:alpham0_1pi_beta0_25pi_phsprt}(d)}
\pointertootherfig{84}{40}{\ref{fig:alpham0_1pi_beta0_25pi_phsprt}(e)}
\end{overpic}
\caption{
Dynamics for $\alpha=-\pi/10, \beta=\pi/4$.
(a): stability diagram. $(\omega,a)$ parameter regions are shaded according to which stable states they contain: Drift (\lbldr, pure rotation around cylinder) and/or locked states (\lblL), librations (\lbllibration, periodic or chaotic), mixed (libration/rotation) oscillation (\lblmixedosc,  periodic or chaotic).
(b): Bifurcation diagram for $\alpha=-\pi/10, \beta=\pi/4, a=-1.9425$ (value of $a$ is marked as \dIndicLine\ in (a)). The diagram shows
stable nodes (\bdSTABLENODE s), saddles, unstable nodes (both \bdSADDLE s)
stable spirals (\bdSTABLEFOCUS s), unstable spirals (\bdUNSTABLEFOCUS s), stable oscillations (local extrema marked in \bdOscExtr) and unstable limit cycles (local extrema marked as \bdUNSTABLELC s).
(c): magnification of (b) (see \dIndicRect\ in (b)). The oscillations (\lblLibrationAndMixedOsc) alternate between periodic and chaotic as $\omega$ increases, and for $\omega>0.78299$ the oscillations are exclusively pure drift states (\lbldr).
(d): magnification of (c) (see \dIndicRect\ in (c)).
The first PD bifurcation (left \bdPD) produces an unstable limit cycle (\bdUNSTABLELC) which is destroyed in a homoclinic bifurcation (HC), and a stable, period-2 limit cycle (\lbllibration) which undergoes a second PD bifurcation (right \bdPD) and then a cascade of PD bifurcations (\textit{not} marked, for readability), leading to stable chaos.
An \lblmixedosc\ state co-existing with the limit cycle is marked in \bdOscExtrCoexist\ for distinguishability.
At $\omega\approx0.6955$, the stable chaotic attractor turns into a stable periodic oscillation.
The maximum Lyapunov exponent $\maxLE$ of the stable oscillatory states is shown beneath.
For further explanations see text. For phase portraits of the oscillations see Fig.\ref{fig:alpham0_1pi_beta0_25pi_phsprt}.
}
\label{fig:alpham0_1pi_beta0_25pi_stabbif}
\end{figure*}
\begin{figure*}
\begin{overpic}[width=.25\linewidth]{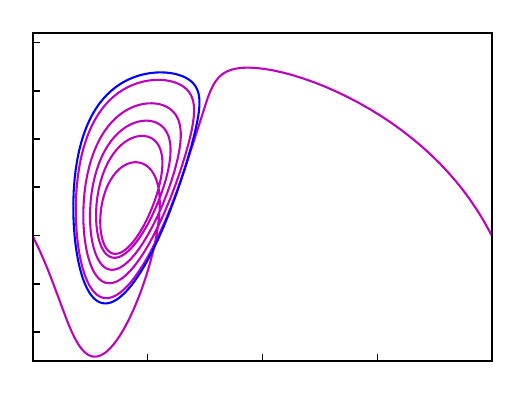}
\pnhdln{$\omega=0.693$,\,\,\,\,\lblLibrationAndMixedOsc}
\pnlblquarter{a}
\lblPhiKapMultimodeDriftChaos
\put(36,11){
\begin{overpic}[width=0.12\linewidth]{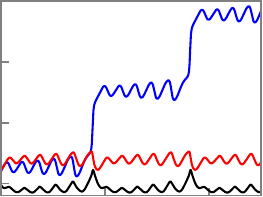}
\put(0,-5)  {\tiny\textcolor{gray}{$0$}}
\put(50,-6)  {\tiny\textcolor{gray}{$t$}}
\put(90,-5) {\tiny\textcolor{gray}{$250$}}
\put(-6,3)  {\tiny\textcolor{gray}{$0$}}
\put(-10,48){\tiny\textcolor{gray}{$10$}}
\put(45,45){\tiny\textcolor{blue}{$\phi$}}
\put(45,21){\tiny\textcolor{red}{$\kappa_{12}$}}
\put(45,8){\tiny\textcolor{black}{$\kappa_{21}$}}
\end{overpic}
    }
\put(50,47){\color{gray}\vector(0,1){14}}
\end{overpic}
\nolinebreak
\begin{overpic}[width=.25\linewidth]{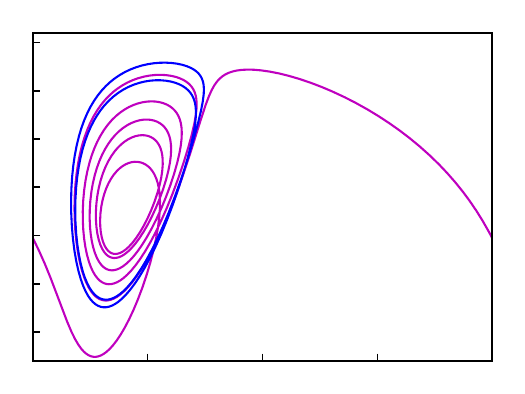}
\pnhdln{$\omega=0.69368$,\,\,\,\,\lblLibrationAndMixedOsc}
\pnlblquarter{b}
\lblPhiKapMultimodeDriftChaos
\end{overpic}
\nolinebreak
\begin{overpic}[width=.25\linewidth]{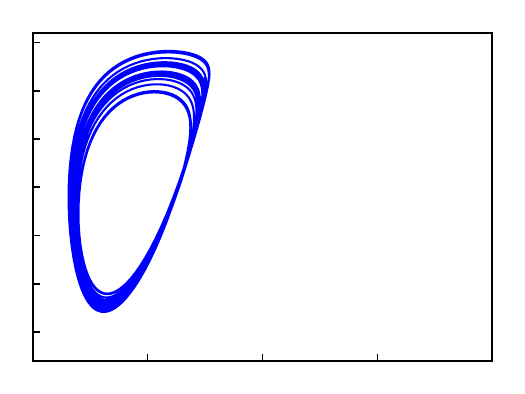}
\pnhdln{$\omega=0.6945$,\,\,\,\,\lbllibration}
\pnlblquarter{c}
\lblPhiKapMultimodeDriftChaos
\end{overpic}
\nolinebreak
\begin{overpic}[width=.25\linewidth]{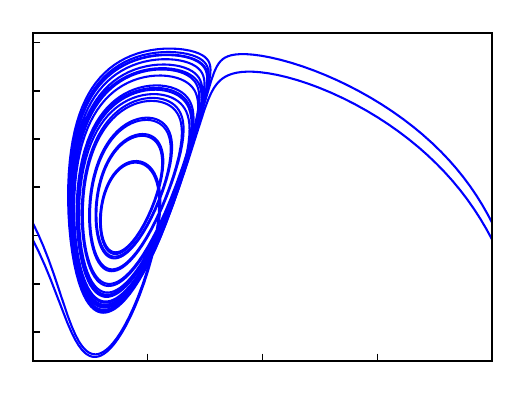}
\pnhdln{$\omega=0.69488$,\,\,\,\,\lblmixedosc}
\pnlblquarter{d}
\lblPhiKapMultimodeDriftChaos
\end{overpic}
\\[4mm]
\begin{overpic}[width=.25\linewidth]{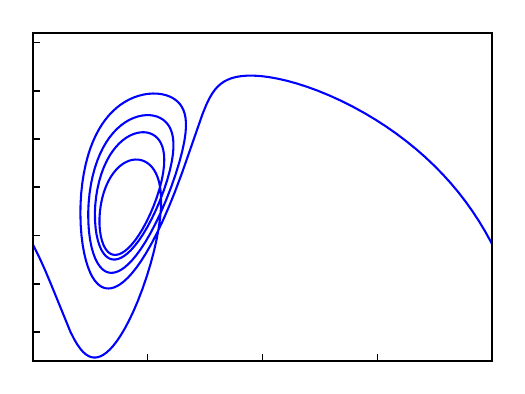}
\pnhdln{$\omega=0.6957$,\,\,\,\,\lblmixedosc}
\lblPhiKapMultimodeDriftChaos
\pnlblquarter{e}
\end{overpic}
\nolinebreak
\begin{overpic}[width=.25\linewidth]{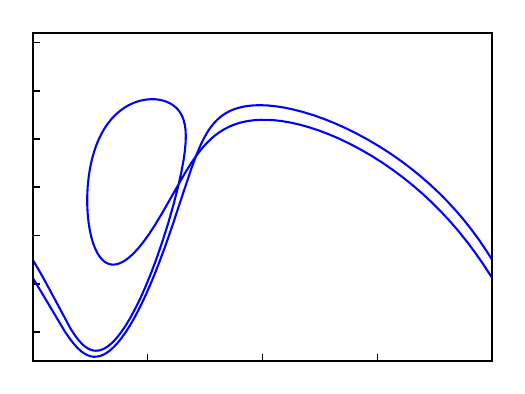}
\pnhdln{$\omega=0.7674$,\,\,\,\,\lblmixedosc}
\pnlblquarter{f}
\lblPhiKapMultimodeDriftChaos
\end{overpic}
\nolinebreak
\begin{overpic}[width=.25\linewidth]{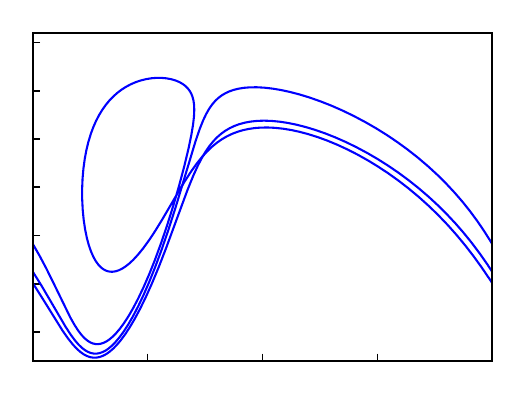}
\pnhdln{$\omega=0.7778$,\,\,\,\,\lblmixedosc}
\pnlblquarter{g}
\lblPhiKapMultimodeDriftChaos
\end{overpic}
\nolinebreak
\begin{overpic}[width=.25\linewidth]{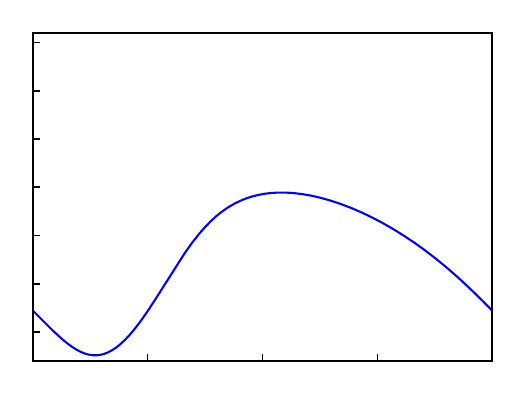}
\pnhdln{$\omega=1$,\,\,\,\,\lbldr}
\pnlblquarter{h}
\lblPhiKapMultimodeDriftChaos
\end{overpic}
\caption{
Regime of supercritical Hopf bifurcations ($a<a_\text{GH}$): phase portraits of stable oscillations for $\alpha=-\pi/10, \beta=\pi/4, a=-1.9425$ and varying $\omega$ (shown as \dIndicLine s in Figs.~\ref{fig:alpham0_1pi_beta0_25pi_stabbif}(b)-(d)).
(a),(b): period-1 (a) or period-2 (b) libration cycle (\lbllibration, \bdOscExtr) co-exists with mixed rotational / librational drift cycle (\lblmixedosc, \bdOscExtrCoexist).
(c): chaotic libration (\lbllibration, \bdOscExtr).
(d): chaotic mixed oscillation (\lblmixedosc), consisting of rotations and librations.
(e): periodic \lblmixedosc\ with 1 rotation and 4 librations (\bdOscExtr).
(f): periodic \lblmixedosc\ with 2 rotations and 1 libration.
(g): periodic \lblmixedosc\ with 3 rotations and 1 libration.
(h): pure rotation drift cycle (\lbldr, \bdOscExtr).
See text for further details.
}
\label{fig:alpham0_1pi_beta0_25pi_phsprt}
\end{figure*}

Second, we consider the stability diagram for $\alpha=-\pi/10$ in Fig.~\ref{fig:alpham0_1pi_beta0_25pi_stabbif}. The resulting dynamics become far more involved when compared to the previously considered cases. The Hopf curve (\sdHB) that emanates from the BT$_1$ point (\sdBT\ in Fig.~\ref{fig:alpham0_1pi_beta0_25pi_stabbif}(a)) contains a Generalized Hopf (GH) point located at ($\omega_\text{GH}$,$a_\text{GH}$) (\sdGH\ in Fig.~\ref{fig:alpham0_1pi_beta0_25pi_stabbif}(a)), which separates the Hopf curve (\sdHB) into a supercritical segment (below GH), and a subcritical segment (above GH).

\subparagraph{Regime of subcritical Hopf bifurcations ($a>a_\text{GH}$).}
The subcritical Hopf curve (\sdHB\ in Fig.~\ref{fig:alpham0_1pi_beta0_25pi_stabbif}(a)) produces unstable (libration) limit cycles, which may undergo a saddle-node-of limit cycles (SNLC) bifurcation (\sdSNLC) before they die in a homoclinic bifurcation (\sdHC). The homoclinic bifurcation curve, which originates in the point BT$_1$ (\sdBT), was found via numerical continuation using MatCont~\cite{Dhooge2008}.
Below the GH point (\sdGH), there are two branches of SNLC bifurcations which meet in a cusp of SNLCs (CPC, \sdCPC). Thus, between the GH and the CPC points, limit cycles produced in the Hopf bifurcation (\sdHB) undergo \textit{two} SNLC bifurcations before dying in the homoclinic bifurcation (\sdHC).
One of the SNLC branches (\sdSNLC) meets the homoclinic bifurcation curve (\sdHC) in a point which we call SLH (\sdSLH).

\subparagraph{Regime of supercritical Hopf bifurcations ($a<a_\text{GH}$).}

The supercritical Hopf bifurcations (\sdHB\ in Fig.~\ref{fig:alpham0_1pi_beta0_25pi_stabbif}(a)) below the GH (\sdGH) produce stable (libration) limit cycles (\lbllibration) which can only exist on the same side of the supercritical Hopf curve as the unstable spirals, i.e., in the \sdLCRL\ between the HB curve (blue) and the HC curve (purple) that destroys the limit cycles.
However, the stable limit cycle can be destroyed / destabilized before reaching the HC bifurcation (\sdHC) in an intriguing bifurcation scenario.  We let $a=-1.9425$ and follow the \dIndicLine\ in the diagram Fig.~\ref{fig:alpham0_1pi_beta0_25pi_stabbif}(a)) while varying $\omega$ and show example trajectories for chosen values of $\omega$ in Fig.~\ref{fig:alpham0_1pi_beta0_25pi_phsprt}.

After the supercritical Hopf bifurcation (right \bdHB\ in Fig.~\ref{fig:alpham0_1pi_beta0_25pi_stabbif}(b)), i.e. for increasing $\omega$, the stable limit cycle \lbllibration\ (\bdOscExtr\ in Figs.~\ref{fig:alpham0_1pi_beta0_25pi_stabbif}(b),(d) and ~\ref{fig:alpham0_1pi_beta0_25pi_phsprt}(a)) undergoes a Period-Doubling (PD) bifurcation (left \bdPD\ in Fig.~\ref{fig:alpham0_1pi_beta0_25pi_stabbif}(d)).
The resulting stable limit cycle (\bdOscExtr\ in Figs.~\ref{fig:alpham0_1pi_beta0_25pi_stabbif}(d) and \ref{fig:alpham0_1pi_beta0_25pi_phsprt}(b)) has period 2. The original, period-1 cycle has become unstable (\bdUNSTABLELC\ in Fig.~\ref{fig:alpham0_1pi_beta0_25pi_stabbif}(d)) in the PD (left \bdPD\ in Fig.~\ref{fig:alpham0_1pi_beta0_25pi_stabbif}(d)); the associated branch is ultimately destroyed in the homoclinic bifurcation (\bdHC\ in Fig.~\ref{fig:alpham0_1pi_beta0_25pi_stabbif}(b) and (c)).
Remarkably, note that the libration cycle \lbllibration\ co-exists with a (rotational) drift limit cycle in the region of period-1 and period-2 cycles, see \bdOscExtrCoexist\ curves in  Fig.~\ref{fig:alpham0_1pi_beta0_25pi_stabbif}(d) and \bdOscExtrCoexist\ trajectories in Fig.~\ref{fig:alpham0_1pi_beta0_25pi_phsprt} (a) and (b). Unlike the previously described \lbldr\ state, these drift cycles consist of both rotations and librations, which is why refer to them as {mixed oscillations} (\lblmixedosc).
While the precise nature of the bifurcation mechanism giving rise to its birth on the left for smaller $\omega$  remains open, inspection of trajectories for larger $\omega$ on the right strongly suggests that this mixed oscillation is destroyed in a collision with the unstable period-1 version of the libration cycle (\bdUNSTABLELC\ in Fig.~\ref{fig:alpham0_1pi_beta0_25pi_stabbif}(d)).

As we increase $\omega$ further, the stable period-2 cycle undergoes a period doubling cascade (PDs are not explicitly marked to improve readability) resulting in a chaotic attractor with $\phi$ bounded, i.e., it  can (still) be seen as a libration \lbllibration\ (\bdOscExtr\ in Figs.~\ref{fig:alpham0_1pi_beta0_25pi_stabbif}(c), (d) and ~\ref{fig:alpham0_1pi_beta0_25pi_phsprt}(c)).
To characterize this chaotic motion we numerically estimated the maximum Lyapunov exponent $\maxLE$, and we found that $\maxLE>0$ (close to zero) throughout the (non-)chaotic ranges of $\omega$ (Fig.~\ref{fig:alpham0_1pi_beta0_25pi_stabbif}(c),(d)).
(Fluctuations in $\maxLE$ while varying $\omega$  are due to the fact that phase points of the original and the perturbed trajectory can lie anywhere on the chaotic attractor by the simulation time $\maxLE$ is computed.)

This chaotic attractor can be seen as a chaotic libration, since $\phi$ does not rotate around the cylinder $\T\times\R$ (Fig.~\ref{fig:alpham0_1pi_beta0_25pi_phsprt}(c)). Remarkably, as $\omega$ is further increased, the chaotic attractor
at some point also includes
oscillations characterized as \textit{rotations}, where $\phi$ revolves around the cylinder  (Fig.~\ref{fig:alpham0_1pi_beta0_25pi_phsprt}(d)) (in a fashion that is reminiscent of \textit{phase slips}).
More precisely, the trajectory librates a number of times, until it escapes and becomes a rotation with drift character, while, however, remaining chaotic in nature. Presumably, this behavior is akin to a mixed mode oscillation seen characteristic of slow-fast systems, which is a reminiscence from low $\epsilon$ (recall that we use a fairly large $\epsilon=0.2$ here).
Eventually, as $\omega$ is increased even further, the cycle ceases to display chaoticity and becomes a periodic oscillation including both librations (of a certain period) and (drift) rotations
(Fig.~\ref{fig:alpham0_1pi_beta0_25pi_phsprt}(e)), which we therefore refer to as \emph{mixed oscillations} (\lblmixedosc; not to be confused with mixed-mode oscillations~\cite{kuehn2015multiple}). Ultimately, as we further increase $\omega$, the librations disappear and the cycle is a purely rotational drift cycle (Fig.~\ref{fig:alpham0_1pi_beta0_25pi_phsprt}(h)).
This is remarkable, since ---  as already noted further above --- a stable mixed oscillation cycle was already present for smaller $\omega$ (\bdOscExtrCoexist\ curves in Figs.~\ref{fig:alpham0_1pi_beta0_25pi_stabbif}(c),(d) and \ref{fig:alpham0_1pi_beta0_25pi_phsprt}(a),(b)), which was destroyed in a bifurcation scenario different from the scenario seen here. Thus, these (rotational/librational) drift  cycles just described are indeed distinct, as they seem to emerge from the libration cycles originally born in the supercritical Hopf bifurcation.
To summarize, stable oscillations encountered are all either
non-drifting libration cycles (\lbllibration, periodic or chaotic),
purely rotational drift states (\lbldr, periodic),
or mixed rotational / librational drift oscillations (\lblmixedosc, periodic or chaotic).

Finally, note that states containing both librations and rotations exist, for $a=-1.9425$, in the range $\omega\in[0.69456, 0.78299)$ (Figs.~\ref{fig:alpham0_1pi_beta0_25pi_stabbif}(c) and ~\ref{fig:alpham0_1pi_beta0_25pi_phsprt}(d)-(g)), where subintervals hosting  periodic cycles alternate with subintervals hosting (aperiodic) chaotic motion. This was found via quasi-continuation of the stable oscillatory states in the interval $\omega\in[0.693, 1.25]$.
Interestingly, the respective number of rotations and librations per period of the oscillation is different for each periodic subinterval (see  Figs.~\ref{fig:alpham0_1pi_beta0_25pi_phsprt}(e)-(g)).
Finally, based on quasi-continuation of the drift states in the direction of $\omega\tz$ for several values of $a$, we find that stable purely rotational drift states (\lbldr) exist only in the \sddr s and \sdLdr s in Fig.~\ref{fig:alpham0_1pi_beta0_25pi_stabbif}(a).
While the \sdLCRL\ \textit{may} contain all aforementioned types of \lbllibration\ and \lblmixedosc\ cycles, none of these types emerges everywhere inside of the \sdLCRL; for readability, we do not distinguish existence regions for each type individually.

\begin{figure*}
\begin{overpic}[width=.25\linewidth]{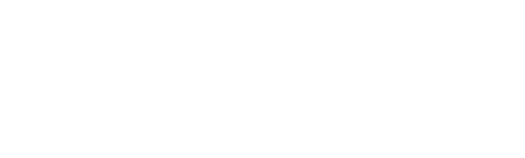}
  \put(0,35){(a)}
\end{overpic}
\nolinebreak
\begin{overpic}[width=.25\linewidth]{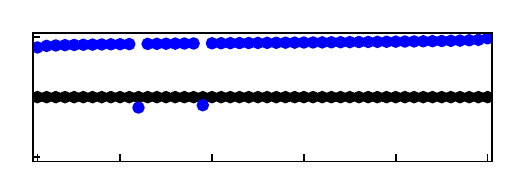}
  \put(0, 35){(b)}
  \put(13,35){$a=1.7,\sigma=0.2$,\,\,\,\,\,\lblAP}
  \put(1,28){$\pi$}
  \put(1,16){$0$}
  \put(-5,6){$-\pi$}
  \put(60,-2){$l$}
  \put(6,0){$1$}
  \put(89,0){$N$}
\put(37,22){\textcolor{blue}{phases $\phi_l$}}
\put(47,9){frequencies $\dotxp{\phi_l}$}
\end{overpic}
\nolinebreak
\begin{overpic}[width=.25\linewidth]{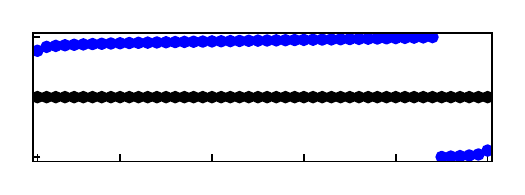}
  \put(0, 35){(c)}
  \put(13,35){$a=1.7,\sigma=0.4$,\,\,\,\,\,\lblL}
  \put(1,28){$\pi$}
  \put(1,16){$0$}
  \put(-5,6){$-\pi$}
  \put(60,-2){$l$}
  \put(6,0){$1$}
  \put(89,0){$N$}
\end{overpic}
\nolinebreak
\begin{overpic}[width=.25\linewidth]{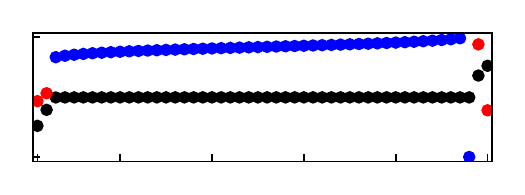}
  \put(0, 35){(d)}
  \put(13,35){$a=1.7,\sigma=0.6$,\,\,\,\,\,\lblpartial}
  \put(1,28){$\pi$}
  \put(1,16){$0$}
  \put(-5,6){$-\pi$}
  \put(60,0){$l$}
  \put(6,0){$1$}
  \put(89,0){$N$}
\end{overpic}
\begin{overpic}[width=.25\linewidth]{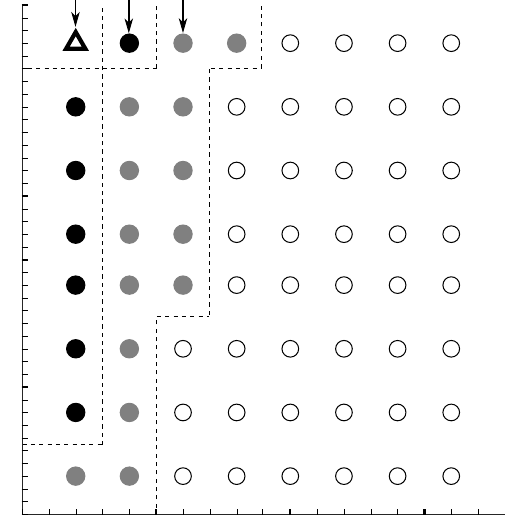}
  \put(10.5,101){(b)}
  \put(21,101){(c)}
  \put(31,101){(d)}
  \put(4,92){\lblAP}
  \put(12.,71){\lblL}
  \put(26,59){\lblpartial}
  \put(58,48){\lbldr}
  \axlblquarter{$\sigma$}{$a$}
  \put(2,-5){$0$}
  \put(76,-5){$1.5$}
  \put(-7,0){$-2$}
  \put(-1,48){$0$}
  \put(-1,97){$2$}
\end{overpic}
\nolinebreak
\begin{overpic}[width=.25\linewidth]{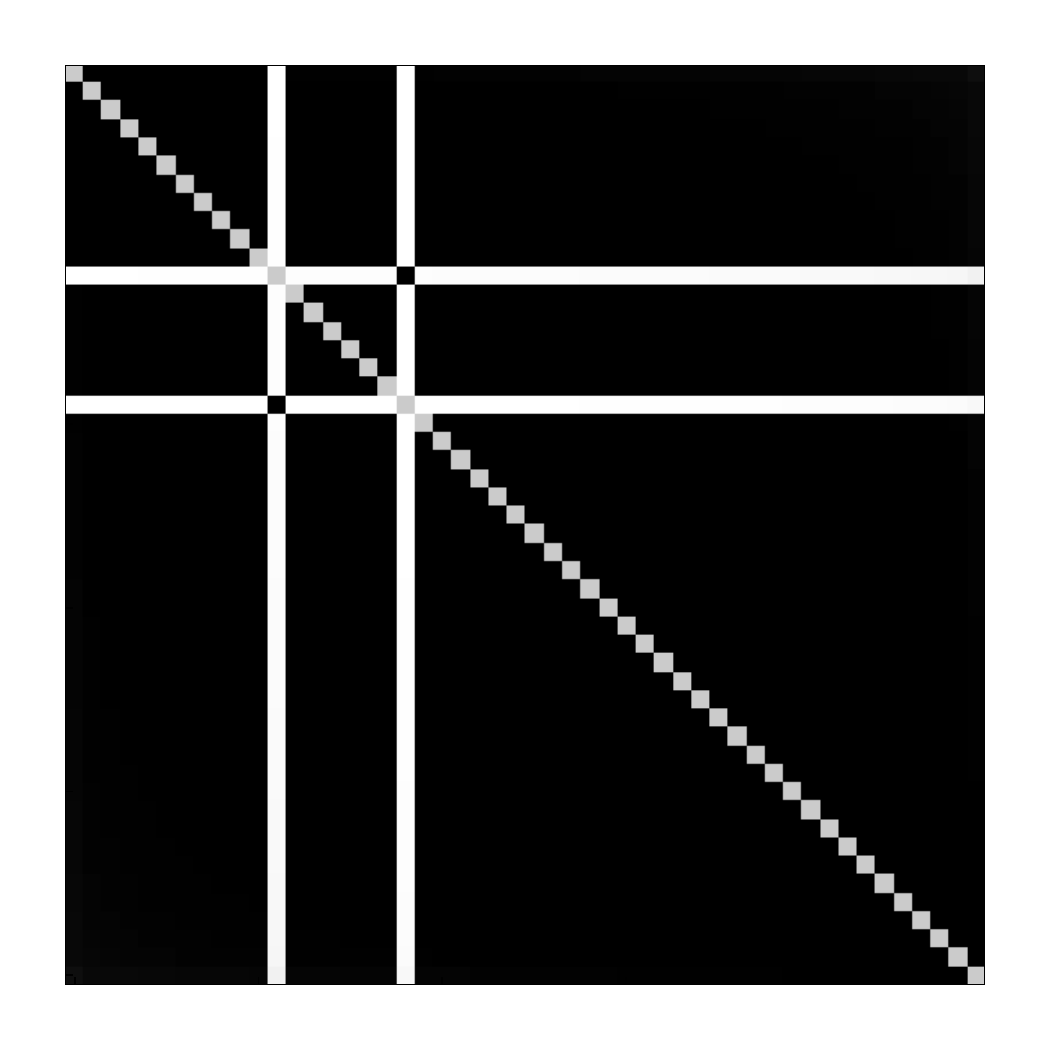}
  \axlblquarter{$m$}{$l$}
  \put(0,89){$1$}
  \put(-1,6){$N$}
  \put(6,0){$1$}
  \put(89,0){$N$}
\end{overpic}
\nolinebreak
\begin{overpic}[width=.25\linewidth]{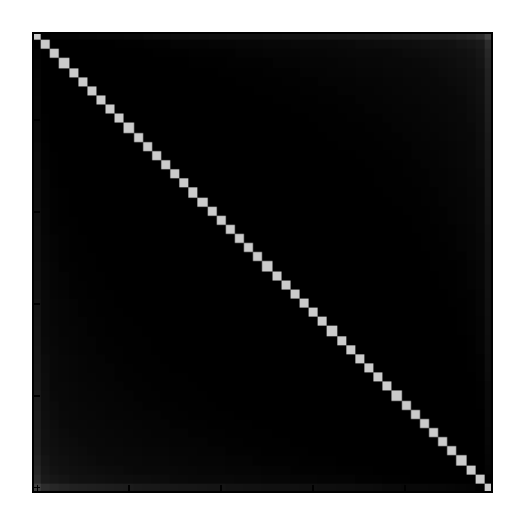}
  \put(0,89){$1$}
  \put(-1,6){$N$}
  \axlblquarter{$m$}{$l$}
  \put(6,0){$1$}
  \put(89,0){$N$}
\end{overpic}
\nolinebreak
\begin{overpic}[width=.25\linewidth]{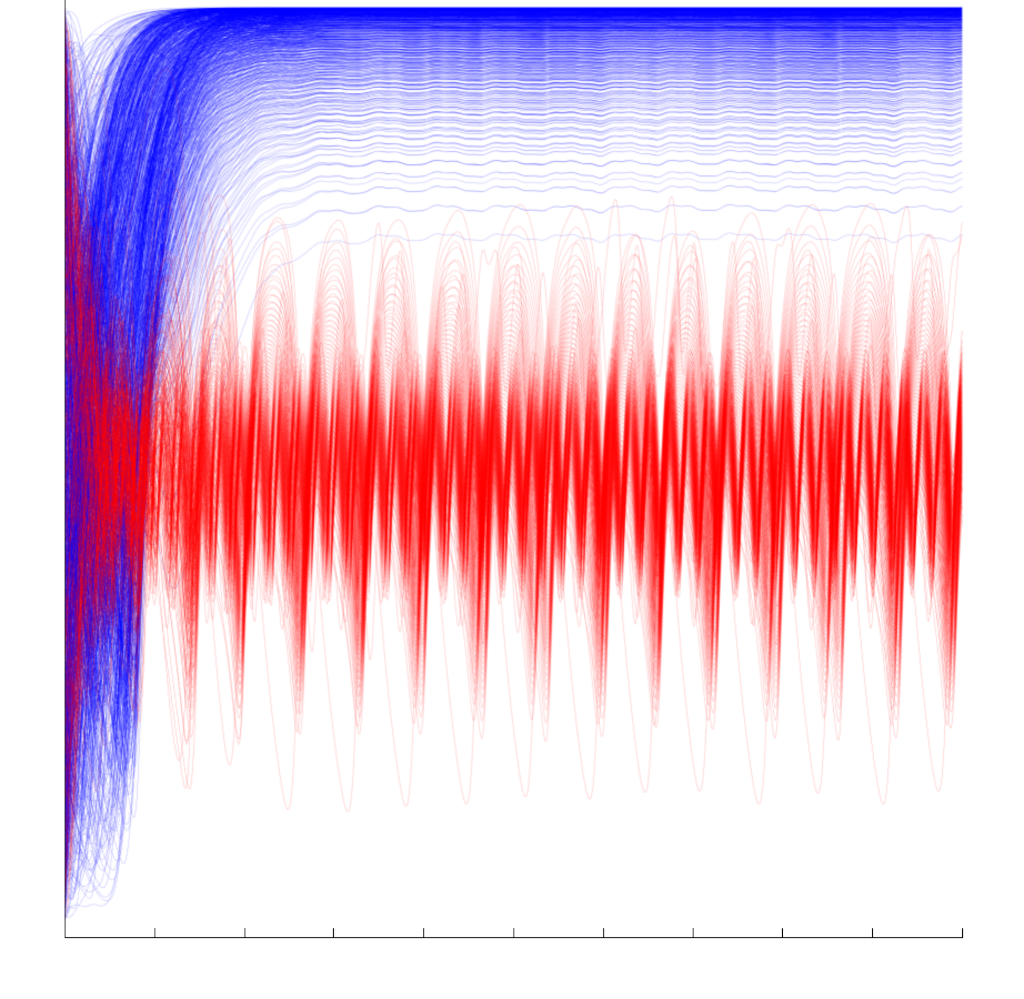}
  \axlblquarter{$t$}{$\kappa_{lm}$}
  \put( 6,  6) {\color{black} \line(1,0){ 88.7}}
  \put( 6.2,  6) {\color{black} \line(0,1){ 91.6}}
  \put(-3,90){$2.7$}
  \put(-4,6){$-0.7$}
  \put(5,0){$0$}
  \put(80,0){$200$}
  \put(42,47){Drifting}
  \put(42,80){Locked}
\end{overpic}
\caption{
Summary of simulation results for $N=50$ oscillators with $\alpha=\beta=0,\epsilon=0.2$.
(a) Stability diagram for varying $a$ and $\sigma$ summarizes the four post-transient  final states  (as observation for one realization only):
antipodal (\lblAP, triangle), frequency-locked (\lblL, black circles), partially synchronous (\lblpartial, gray circles), drift (\lbldr, empty circles).
Panels (b)-(d) display the behavior observed for three distinct parameter values (highlighted with \dIndicRect s in (a)), i.e.,
\lblAP\ for $a=1.7,\sigma=0.2$ (b), \lblL\ for $a=1.7,\sigma=0.4$ (c), and \lblpartial\  $a=1.7,\sigma=0.6$ (d), respectively.
The top row in the panels displays phases (blue/red dots for locked/drifting oscillators), $\phi_l$, as well as dynamic frequencies (black dots), \highlight{$\dotxp{\phi_l}$ in Eqs.~\eqref{eq:reducedmodel},} at the final simulation time $T$.
The bottom row in panels (b) and (c) features the $\kappa$ matrices at final time $T=1500$, where white and black pixels correspond to $1-a=-0.7$ and $1+a=2.7$, respectively (both matrices have zero diagonal shown in gray). \highlight{Note that the black/white regions are not uniformly valued, but heterogeneous values (see text).}
The bottom row in panel (d) presents the time evolution of $\kappa_{lm}(t)$ during $0\leq t\leq T=200$ (transients have subsided for $t\leq 50$).}
\label{fig:N50}
\end{figure*}
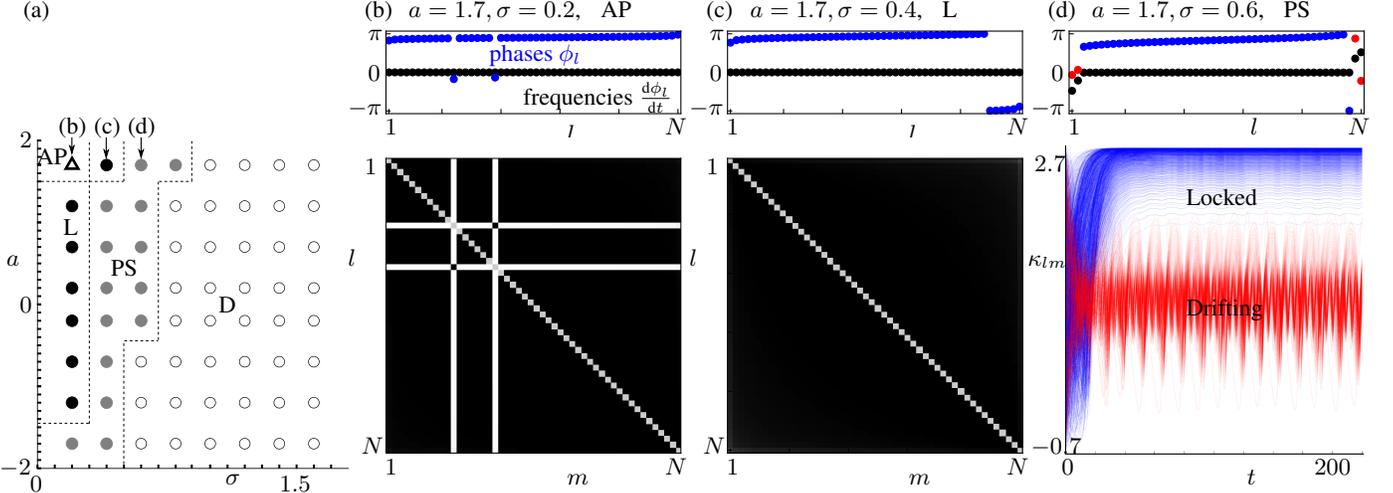

\section{Simulations for $N=50$ oscillators}\label{sec:N50}
In Sec.~\ref{sec:analysis}, we have studied in detail the behavior of the system~\eqref{eq:reducedmodel} for $N=2$ oscillators. The question abounds whether some of the \highlight{behavior observed for} only $N=2$ oscillators carries over to a larger version of the system with $N=50$ oscillators. In order to address this  question we restrict our attention to the simplest case with parameters $\alpha=\beta=0,\epsilon=0.2$, (see Figs.~\ref{fig:2Dbif},~\ref{fig:2Dstab} for the results obtained for $N=2$).
For two oscillators we could tune the frequency mismatch, $\omega$; for the many oscillator system, we instead let the intrinsic frequencies $\omega_l$ follow a normal distribution with zero mean and standard deviation $\sigma$. Furthermore, instead of drawing these frequencies randomly, we constructed them via a an equidistant set of points mapped to the interval $[-3\sigma, 3\sigma]$ via the inverse error function. This symmetrization eliminates potential random effects that \highlight{due to finite size effects} might otherwise obscure the dynamic effects we are interested in, such as to render comparison to the case of $N=2$ more immediate.

We study the dynamic asymptotic behavior of this system on a grid in $(\sigma,a)$ parameter space,
the results of which are summarized in Fig.~\ref{fig:N50}(a). \highlight{We may distinguish four states:  the (frequency-)locked state (\lblL) and the antipodal state (\lblAP), two frequency-locked states where all $\d\phi_l/\d t ,\d{\kappa_{lm}}/d t$ tend to zero after some transient,; a drift (\lbldr) state with all oscillator phases drifting a part; and the partial \highlight{synchrony} state (\lblpartial) which can be seen as a mixture of both \lblL\ and \lbldr\ states, in the sense that the $N$ oscillators are split into a frequency-locked and a drifting subset. We describe these states in the order of appearance as we increase $\sigma$ for $a=1.7$ as shown in Fig.~\ref{fig:N50}.}

For the \emph{antipodal/anti-phase state} (\lblAP) (Fig.~\ref{fig:N50}(b)), the phases of oscillators split into two groups inside which phases differ only very little from one another, but in between the groups phases differ by about $\pi$, i.e., the two oscillator groups are in \highlight{an anti-phase (antipodal) configuration with respect to one another} on the phase circle;  $\d{\phi_l}/\d t,\d{\kappa_{lm}}/\d t$ tend to zero asymptotically in time. This special configuration of the phases also impacts the coupling on the network: $\kappa_{lm}$ that represent couplings between oscillator pairs residing in the same population (distinct populations) assume values close to the maximum possible value $1+a$ (the minimum possible value $1-a$), see black (white) $\kappa_{lm}$ values in Fig.~\ref{fig:N50}(b). Note that these values for $\kappa_{lm}$ are heterogeneously distributed near these values as they accommodate for the non-identical phase configuration (the figure compresses the variation due to the chosen gray scale).
Note that the antipodal state occurs when the distribution of $\omega_l$ is sufficiently narrow and the adaptivity $a>0$ is sufficiently strong. This state may be interpreted as a corollary to the \lblLL\ co-existence observed for $N=2,\alpha=\beta=0,a\gg0,|\omega|\ll1$ (see Fig.~\ref{fig:2Dbif}(a)), where the two oscillators either occupy a phase configuration with 0 or $\pi$  difference; here, on the other hand, oscillators form groups \highlight{of varying size} that mutually adhere to either of the two configurations. Note that a similar correspondence applies to the coupling strengths both at the upper and the lower end of the applicable $\kappa$ range.

For the \emph{frequency-locked state} \lblL\ (Fig.~\ref{fig:N50}(c)), all phases are close to one another. This configuration implies that the $\kappa_{lm}$ are close to $1+a$ (and bounded by that value from above). \highlight{Note that individual coupling strengths $\kappa_{lm}$ for $l$ and $m$ are not identical but heterogeneously distributed: the coupling strengths have adapted to accommodate for the non-identical phase configuration. (In Fig.~\ref{fig:N50}(b) and (c) the gray scale for the values of $\kappa_{lm}$ ranges from $1-a$ to $1+a$  so that the distributed values cannot easily be distinguished}. Topologically, this essentially corresponds to an all-to-all network. This state is thus interpreted as the corollary of the frequency-locked state \lblL\ for $N=2$, see Fig.~\ref{fig:2Dbif}(b) for small $|\omega|$.

For the \emph{partially synchronous state} (\lblpartial) (in Fig.~\ref{fig:N50}(d)),  oscillators split into two groups where one is locked (blue dots for phases), while the other contains drifters (red dots for phases). Correspondingly, coupling strengths $\kappa_{lm}$ where $l$ and/or $m$ can be associated with drifting oscillators and are thus oscillatory (red curves in Fig.~\ref{fig:N50}(d)).
By contrast, the locked oscillators (blue dots in Fig.~\ref{fig:N50}(d)) display vanishingly small dynamic frequencies, thus the $\kappa_{lm}$ for which neither $l$ nor $m$ are drifting oscillate close to $1+a$ and with very small amplitude (blue time traces in Fig.~\ref{fig:N50}(d)). 
\highlight{ We note that the correspondence to states observed for $N=2$ is less obvious; however,
 an extension of the D/L bistable region for $N=2$ oscillators to larger systems could give space to accommodate for configurations where a group of oscillators populate a lock state while the remaining oscillators adhere to drifting oscillations.
}

For the \emph{drifting state} (\lbldr), oscillator phases are \emph{incoherent} and drift apart from each other in unlocked motion while the coupling strengths $\kappa_{lm}$ appear to attain a (quasi-)stationary distribution centered around 1 (not shown). We interpret this state as the corollary to the \lbldr\ state for $N=2$.

Just as for the $N=2$ state, the locked state \lblL\ occurs for both positive and negative $a$. For $N=50$ and $a<0$ with $\sigma$ small, we observe locked states (corresponding to stable foci for $N=2$) whose $\kappa_{lm}$ are mostly close to the minimal value $1-a$, in analogy to the case of $N=2$ oscillators (see Fig.~\ref{fig:2Dbif}(c)).

For the $N=2$ system, we found that (co)existence of one (\lblL) or two (\lblLL) stable locked state(s) is guaranteed for $|a|$ sufficiently large. Interestingly, Fig.~\ref{fig:N50}(a) suggests that there is a stable locked state for $a>0$ sufficiently large; yet, for $a<0$ it appears as if the region of stable locked states is diminished in favor of partially synchronous\highlight{, and, even drift states (incoherence). Similarly, we observed for $N=2$ and $a<0$  that the locking region is diminished when compared to $a>0$, which is due to a subcritical Hopf bifurcation that destabilizing the frequency-locked branch (albeit not as much as strongly as for $N=50$) --- we hypothesize that this effect is more pronounced for larger oscillator numbers (and ensuing bifurcations). It would be interesting to investigate this aspect in future research.}

\section{Discussion}\label{sec:discuss}

The model in Eqs.~\eqref{eq:reducedmodel} is a Kuramoto-Sakaguchi model with phase-lag $\alpha$ and coupling strengths that adapt according to a learning rule with adaptivity strength $a$ and adaptation shift $\beta$. \highlight{The parameter $a$ allows to tune away from the limit of the non-adaptive classical Kuramoto model with stationary coupling ($a=0$) and systematically investigate the effect of adaptivity on the collective dynamics.}

Our bifurcation analysis concerns the case of $N=2$ oscillators, for which we --- chiefly --- may distinguish three paradigms of adaptivity: (i) the non-adaptive limit with stationary coupling ($a=0$); (ii) \highlight{symmetric adaptation} ($\beta=0$ or $\beta=\pi$); and (iii) asymmetric adaptation (arbitrary $\beta$).
The non-adaptive limit ($a=0$) trivially reduces to the classical Kuramoto model,
where, for frequency locking to occur, the frequency mismatch must be smaller than the coupling strength between oscillators.

{Considering the case of \highlight{asymmetric} coupling ($\beta=0$), we first observe that deviating from the non-adaptive limit ($a=0$) with non-zero adaptivity ($a\neq0$) leads to an overall larger locking region (\lblL), i.e., a larger frequency mismatch is required to break locking. It is interesting to note that in the context of forced oscillations, the mode-locking region corresponds to an Arnol'd tongue; indeed, the strength of adaptivity could be related to a forcing strength acting on one of the oscillators, at least in the limit of slow adaptation ($\epsilon\rightarrow 0$).
Furthermore, smaller frequency mismatch $|\omega|$ and sufficiently large adaptivity $a$ allow for bistable regions \lblLL\ where two frequency locked modes exist, i.e., anti-phase (antipodal) configurations co-exist in addition to in-phase configurations 
\highlight{--- indeed, considering stationary coupling in, e.g., Eq.~\eqref{eq:governing2DRescal} effectively results in a bi-harmonic coupling function which may lead to such bi-stability~\cite{Ashwin2015weak}}. 
For larger $|\omega|$ and $a\neq0$, a further bistability region \lblLdr\ appears  where locked states co-exist with drift cycles (rotations). While this general picture prevails for both positive and negative adaptivity, the situation is slightly more complicated in the region with $a<0$, where  additional bifurcations diminish the width of the various (bi-)stable regions (\lblL, \lblLL\ and \lblLdr).  Drift cycles correspond to rotational limit cycles around the cylindrical phase space, whereas Hopf bifurcations give rise to librational limit cycles; however, in the case of \highlight{symmetric} coupling ($\beta=0,\pi$), these cycles remain always unstable.

We found that similar bifurcations scenarios occur for \highlight{networks with asymmetric coupling} for $\beta\neq0$  and $a>0$; but, even more intriguing dynamics are possible for $a<0$ and $\beta=\pi/4,\alpha=-\pi/10$. In short, a generalized Hopf bifurcation (GH) may be present, around  which a complex structure of bifurcations is organized (including sub-/supercritical Hopf, saddle-node of limit cycles (SNLC), cusp of cycles (CPC), a homoclinic and SLH bifurcations, similar to a scenario seen for a system of Theta neurons, see Ref.~\onlinecite{Juttner2021}); in particular, \highlight{for $a<0$ the GH may give rise to a supercritical Hopf bifurcation leading to stable librational limit cycles.}
These libration cycles may undergo a further period-doubling cascade to chaos. Moreover, remarkably, period-1 and period-2 \highlight{libration} cycles may co-exist with a rotational drift cycle characterized by a non-trivial winding number; \highlight{this rotational} cycle appears to be destroyed in a collision with the unstable period-1 cycle generated by the first period-doubling bifurcation, before the stable  libration cycle becomes chaotic. However, the period-doubling cascade of these librations features a surprise. As $\omega$ increases, after the librations have become chaotic, parameter regions exist where the librations are altered into a mixed oscillation, which combines features of both libration and rotation. The winding number of the libration gradually changes, and mixed oscillations alternate between chaotic and periodic, until for large $\omega$ one finds a regular period-1 rotational drift cycle.

One may expect that the analysis of the $N=2$ oscillator system might be able to capture certain aspects of the dynamic behavior seen in larger systems. Our simulations for \highlight{adaptive networks with symmetric coupling} ($\beta=0$) with $N=50$ revealed that such a correspondence exists, albeit with some limitations. Firstly, considering $a>0$, we find a relatively good correspondence between antipodal/anti-phase (\lblAP), locked (\lblL), and drifting/incoherent states (\lbldr) seen for $N=50$ with the \highlight{bistable locking region (\lblLL), the locking region (\lblL), and drifting region (\lbldr) seen for $N=2$; the partially synchronous (\lblpartial) state for $N=50$ may be related to the \lblLdr\ region for $N=2$}. \highlight{Secondly, increasing strength of adaptivity $|a|$ leads to a widening of the locking regions for $N=2$; this effect appears to be present for $N=50$ only when $a>0$, but not for $a<0$. }

We point out similarities between the system studied here and systems subject to previous research. 
These systems are contained as special cases in our model \eqref{eq:reducedmodel}. The relationships are more obvious when considering the unscaled version \highlight{of our adaptive model} 
\eqref{eq:adaptationrule}: Letting $a_0=0$ with $\alpha=\beta=0$, we recover the model Eqs.~(3) and (4) studied by Seliger {\it et al.}, Ref.~\onlinecite{Seliger2002plasticity}. 
Similarly, if we set $a_0=0$ with $\omega_1=\ldots=\omega_N=0, a_1=1$, we recover the model by Berner {\it et al.}~\cite{Berner2019}. Our adaptation rule generalizes the model in Ref.~\onlinecite{Seliger2002plasticity} by inclusion of
an adaptation shift $\beta\neq 0$ (see Ref.~\onlinecite{Berner2019}); but in contrast to  Ref.~\onlinecite{Berner2019}}, we introduced a nontrivial adaptation offset $a_0$ and strength of  adaptation $a_1$. This choice allowed us to systematically deviate from the classical Kuramoto-Sakaguchi model with stationary coupling, and thereby, to address the question how varying levels of adaptivity impacts the dynamics of the network.
Furthermore, unlike Ref.~\onlinecite{Berner2019}, we allow for non-identical frequencies ($\omega\neq 0$ and $\sigma\neq 0$, respectively) which breaks symmetries in the system --- thus, this enables us to study the width of the locking region. We note that the choice $a_0=1$ does not allow us to retrieve the model in Ref.~\onlinecite{Seliger2002plasticity} since $a_0\tz$ constitutes a singular limit.

Nevertheless, some of the basic dynamic behavior, such as the presence of frequency locking or antipodal (antiphasic) cluster states seen in Ref.~\onlinecite{Berner2019}, naturally carry over to our system as long as intrinsic frequencies are kept relatively close to each other.
Vice versa, breaking the symmetry via disordered frequencies ($\omega\neq 0$) opens the door for bifurcations resulting in novel states, such as small amplitude librations and the associated period-doubling cascade reported here.
A recent study~\cite{Thiele2021asymmetric} carries out an analysis  of our system with $a_0=0$ and $a_0=1$ in the slowly adapting limit ($\epsilon\tz$); however, this analysis separates the dynamics of the coupling strengths into a planar problem, and thus, the resulting dynamics cannot display the transition to chaos or the mixed oscillations containing both librations and rotations that we found here. \highlight{Finally, we note that Kasatkin and Nekorkin~\cite{Kasatkin2016} studied the same  oscillator system with $N=2$ as in this study, except that frequencies are identical ($\omega=0$);  in this case, these authors also observed chaotic dynamics.}

\highlight{
Our simulations for $N=50$ revealed that oscillators may split into two groups which experience distinct strong or weak coupling, depending on the specific phase configuration that the oscillators assume. This phenomenon is particularly prominent for the states we called \lblAP\ and  \lblpartial. WE note that several studies addressed the effect of non-uniform, but \emph{time-independent} coupling on oscillator networks. 
In order to understand the occurrence of chimera states, past research performed bifurcation analysis for $N=4$ oscillators~\cite{Burylko2022symmetry} and for larger $N$, or in the continuum limit ($N\to\infty$)~\cite{Abrams2008,Panaggio2016chimera,Martens2016,HongStrogatz2011,Bick2018}. 
However, drawing analogies to our model is not straight forward. Firstly, the configurations seen for the \lblAP\ or \lblpartial\ state in Fig.~\ref{fig:N50} display groups with different oscillator numbers, but aforementioned studies consider oscillator groups of equal size.  Secondly, the adaptive capability of the systems renders the dynamics intrinsically more complicated. For instance, the adaptive rule in Eq.~\eqref{eq:reducedmodel}b) may give rise to bi-harmonic coupling function (depending on parameter choice) and thus to bi-stability of configurations, see Fig.~\ref{fig:2Dbif}, a feature that is due to higher harmonic interaction~\cite{Ashwin2015weak} (This becomes evident, e.g., by assuming a stationary solution for \eqref{eq:governing2DRescal}). Moreover, the adaptivity may stabilize certain non-uniform coupling configuration, while models without adaption simply enforces the non-uniform coupling. This is a fundamental difference --- as we have seen, adaptive dynamics generates nontrivial dynamic behavior via bifurcations that are absent in the non-adaptive limit (Sec.~\ref{sec:Dyn1D}).
There exist other models with bimodal frequency distributions where the oscillator population may effectively split into two groups adhering to distinct dynamic behavior~\cite{Kuramoto1984,Crawford1994,Martens2009,pazo2009existence,pietras2016equivalence}, and even models that combine bimodally distributed frequencies with non-uniform coupling strength ~\cite{HongMartens2021}; however, how to correlate the observed states with the behavior observed for these systems seems less obvious.
}

We plan to further study adaptive Kuramoto-Sakaguchi oscillator networks with large oscillator numbers. To this end, a variety of questions are interesting to pursue. For instance, it will be interesting to see if similar complex dynamics, such as the period-doubling cascade seen for $N=2$ with \highlight{asymmetric coupling} ($\beta\neq0,\pi$; see Sec.~\ref{sec:Dyn3D}), persists for larger systems.
Moreover, developing a (finite) mean-field theory as well as a continuum limit description for Eqs.~\eqref{eq:reducedmodel}, e.g., using graphons/graphops and other methods~\cite{Gkogkas2022,Gkogkas2022mean,Gkogkas2021continuum,Kuehn2021vlasov,Fialkowski2023,Bick2021mean} would be very useful to study the dynamics of Eqs.~\eqref{eq:reducedmodel}, but remains a formidable challenge. Finally, it would be interesting to study and relate adaptive oscillator networks to more realistic settings such as those that occur in a biological, neuro-scientific or experimental context~\cite{Duchet2022,Calugaru2020,MartensThutupalli2013}. These represent formidable open problems for  future research.

\section{Data availability}
Data sharing is not applicable to this article as no new data were created or analyzed in this study.

\section{Acknowledgements}
We thank M.P. Sørensen, C. Bick, S. Yanchuk, and R. Cestnik for useful discussions and comments throughout this study.
We acknowledge the DTU International Graduate School for support via the EU-COFUND project ``Synchronization in Co-Evolutionary Network Dynamics (SEND)''.

\appendix

\section{Symmetries}\label{app:symm}
Eqs.~\eqref{eq:governingRescal} exhibit various symmetries that we survey here.
The following symmetries allow us to reduce the $\alpha,\beta$ parameter range:
\begin{align}
  (a,\beta) &\mapsto (-a,\beta+\pi)\label{eq:abeta}\\
  (a,\alpha,\phi) &\mapsto (-a,\alpha+\pi,\phi+\pi) \label{eq:aalphaphi}\\
  (\alpha,\beta,\kappa_{12},\kappa_{21}) &\mapsto (-\alpha,-\beta,\kappa_{21},\kappa_{12})\label{eq:minusalphabeta}
\end{align}
Moreover, the following symmetries about the $\omega,a$ axes are visible in all stability diagrams:
\begin{align}
(\phi,\omega,\kappa_{12},\kappa_{21}) &\mapsto (-\phi,-\omega,\kappa_{21},\kappa_{12})\label{eq:swapnames}\
\end{align}
As mentioned in the text, this symmetry defined leaves equilibria of Eqs.~\eqref{eq:governingRescal}, their types and their stabilities unchanged.
Similarly, the symmetry 
\begin{align}
(a,\phi,\kappa_{12},\kappa_{21}) &\mapsto (-a,-\phi+\pi,\kappa_{21},\kappa_{12}) \label{eq:minusaphiinvshift}
\end{align}
preserves equilibria of Eqs.~\eqref{eq:governingRescal}, but the stabilities and bifurcations of these equilibria require a further discussion.
For this, we look at the eigenvalues of the Jacobian specified in \eqref{eq:Jac3Deval}. Clearly, the SN condition $0\neq\mu^2=\delta$ is equivalent to
  \begin{align}
    0=aB-A\label{eq:SNcondaAB}.
  \end{align}
If condition \eqref{eq:SNcondaAB} is fulfilled, then it is still fulfilled after applying the symmetry, since the symmetry leaves $B$ unchanged and only inverts the sign of $a$ and $A$.
Looking at Hopf bifurcations, $\mu=0$ in \eqref{eq:Jac3Dmu} is a necessary condition for a Hopf point. If $\mu=0$ is fulfilled, it is clearly not fulfilled after the symmetry is applied, since the symmetry inverts the sign of $A$. 

\highlight{
In summary, symmetry \eqref{eq:swapnames} preserves equilibria, their stabilities, as well as all bifurcations since we are merely renaming the oscillators; the symmetry \eqref{eq:minusaphiinvshift} preserves SN and cusp bifurcations, but not Hopf bifurcations.
Note that these two symmetries are evident in all stability diagrams shown in this article, i.e.,  Figs.~\ref{fig:2Dstab},~\ref{fig:stabbeta0_5pi_alpham0_1pi},~\ref{fig:stabbeta0_25pi_alpha0_1pi}, and~\ref{fig:alpham0_1pi_beta0_25pi_stabbif}(a).
}

Finally, the following symmetries are not exploited in the text but mentioned here for the sake of completeness:
\begin{align}
(\alpha,\beta,\phi) &\mapsto (\alpha+\pi,\beta+\pi,\phi+\pi) \label{eq:shiftpi},\\
(\alpha,\omega,\epsilon,t) &\mapsto (\alpha+\pi,-\omega,-\epsilon,-t)\label{eq:timerev},\\
(a,\omega,\epsilon,\phi,t) &\mapsto (-a,-\omega,-\epsilon,\phi+\pi,-t)\label{eq:timerevnoalphabeta},\\
(\alpha,\omega,\epsilon,t) &\mapsto (\alpha+\pi,-\omega,-\epsilon,-t). \label{eq:ao}
\end{align}

\bibliographystyle{unsrt}

\end{document}